\documentclass[useAMS,usenatbib]{mn2e}

\usepackage{graphicx}
\usepackage{bm}
\usepackage{natbib}
\usepackage{subfig}
\usepackage{subfloat}

\newcommand{\be}{\begin{equation}}
\newcommand{\ee}{\end{equation}}

\newcommand{\ba}{\begin{eqnarray}}
\newcommand{\ea}{\end{eqnarray}}

\title[Hall equilibria]{Hall equilibria with toroidal and poloidal fields: application to neutron stars}

\author[K.N.~Gourgouliatos et al. ]{\parbox{\textwidth}{K.N.~Gourgouliatos\thanks{E-mail: kostasg@physics.mcgill.ca}$^{1,2}$, A.~Cumming$^{1}$, A. Reisenegger$^{3}$, C. Armaza$^{3}$, M.~Lyutikov$^{4,5}$ and J.~A.~Valdivia$^{6}$ }\vspace{0.4cm}\\
\parbox{\textwidth}{ $^{1}$ Department of Physics, McGill University, 3600 rue University, Montr\`eal, Qu\'ebec H3A 2T8, Canada \\
 $^{2}$ Centre de Recherche en Astrophysique du Qu\'ebec Fellow\\
 $^{3}$  Instituto de Astrof\'isica, Facultad de F\'isica, Pontificia Universidad Cat\'olica de Chile, Av. Vicu\~na Mackenna 4860, 782-0436 Macul, Santiago, Chile \\
$^{4}$ Department of Physics, Purdue University, 525 Northwestern Avenue, West Lafayette, IN 47907-2036, USA\\
$ ^{5}$ The Canadian Institute for Theoretical Astrophysics, University of Toronto,  60 St. George Street Toronto, Ontario, M5S 3H8 Canada\\
$^{6}$ Departamento de F\'isica, Facultad de Ciencias, Universidad de Chile, Casilla 653, Santiago, Chile } }

\begin{document}

\date{Accepted -. Received -; in original form -}
\pagerange{\pageref{firstpage}--\pageref{lastpage}} \pubyear{-}
\maketitle

\label{firstpage}

\begin{abstract}
We present solutions for Hall equilibria applicable to neutron star crusts. Such magnetic configurations satisfy a Grad-Shafranov-type equation, which is solved analytically and numerically. The solutions presented cover a variety of configurations, from purely poloidal fields connected to an external dipole to poloidal-toroidal fields connected to an external vacuum field, or fully confined within the star. We find that a dipole external field should be supported by a uniformly rotating electron fluid. The energy of the toroidal magnetic field is generally found to be a few percent of the total magnetic field energy for the fields with an external component. We discuss the evolution due to Ohmic dissipation which leads to slowing down of the electron fluid. We also find that the transition from an MHD equilibrium to a state governed by Hall effect, generates spontaneously an additional toroidal field in regions where the electron fraction changes.  
\end{abstract} 


\section{Introduction}

Neutron stars contain strong magnetic fields that are known to be steady or evolve very slowly with time. Their external field, responsible for the spin-down of pulsars and magnetospheric activity, is anchored in the crust of the neutron star and has been extensively studied as a force-free or an MHD equilibrium, {\it i.e.}~\cite{GoldreichJulian:1969, Sturrock:1971, Contopoulos:1999, Spitkovsky:2006, Tchekhovskoy:2012}. Evidence from pulsars and quiescent magnetars suggests that the field is in stable equilibrium and small changes in the magnetosphere are radiated away with the exception of some rare giant flares in magnetars \citep{Mazets:1979, Hurley:1999, Palmer:2005} where the magnetospheric field suffers a major rearrangment. In general, any long-term evolution of the magnetospheric field will be because of changes of the magnetic field emanating from the neutron star. Thus, it is important to find solutions of stable fields in the crust to maintain the stable external fields observed. 

The crust can be thought of as a conducting crystal lattice where electrons are free to move and advect the magnetic field.  Except for magnetars, Lorentz forces are weak, and they are balanced by elastic forces from the lattice. Because of this, force-free or MHD equilibria are not applicable. Since only electrons are able to move, the currents are carried by electrons and the evolution of the magnetic field is determined by a combination of the Hall effect and Ohmic decay \citep{Jones:1988, Urpin:1991, GoldreichReisenegger:1992}.

Studies of magnetic field evolution in neutron star crusts have tended to focus on the natural approach of starting with an initial field and following its evolution under the joint action of the Hall effect and Ohmic decay (e.g.~\citealt{Urpin:1991, PonsGeppert:2007, PonsGeppert:2010, Kojima:2012}). Despite its conceptual simplicity, this method requires a knowledge of the field structure at the moment of solidification of the crust. As the formation of a neutron star and its crust involves gravitational collapse and phase transitions, the resulting field will have only a limited resemblance to the progenitor's MHD equilibrium field, which can only be approximately guessed from simulations. Some basic properties, such as total magnetic flux and helicity will be conserved and hint at a starting point, but there is considerable uncertainty in the initial condition for the magnetic field in the crust.

We shall take the reverse approach by looking for long-term equilibrium solutions of the magnetic field under the action of the Hall effect. \cite{Cumming:2004} pointed out that a purely poloidal dipole magnetic field would not evolve under the Hall effect if the toroidal currents were distributed so that the electron fluid rotates rigidly, $J\propto n_{\rm e} r \sin \theta$. The magnetic field lines, advected by the electron fluid, remain unchanged if the electrons are rigidly rotating. The existence of steady solutions under the Hall effect is interesting for a number of reasons. It suggests that accelerated Ohmic decay driven by Hall evolution may not be able to completely dissipate the magnetic energy and could instead saturate. Such a configuration might then be expected to represent the long term configuration of the magnetic field after several Hall-times. Indeed, recent axisymmetric simulations of crustal fields show that the initial evolution due to the Hall effect saturates \citep{PonsGeppert:2007, Kojima:2012}, which provides additional motivation to study steady-state solutions.

In this paper, we show that there is a family of axially symmetric Hall equilibria with mixed poloidal and toroidal fields inside the star and a poloidal field outside the star. These solutions generalize the rigidly rotating, purely poloidal equilibrium found by \cite{Cumming:2004} to include toroidal fields. This is important because a toroidal field is an essential component for the stability of an MHD structure \citep{Mestel:1956, Prendergast:1956, Woltjer:1958, Markey:1973, Wright:1973, FlowersRudermann:1977, BraithwaiteSpruit:2004, Spruit:2008, MarchantReisenegger:2011, Braithwaite:2009, Braithwaite:2012}, so that even if the stability of a Hall equilibrium is possible with a purely poloidal field, the progenitor field will have a non-zero helicity and thus the toroidal field cannot be entirely dissipated. Toroidal fields are also essential for the magnetar energy budget \citep{Thompson:1995}.  

We find axisymmetric Hall equilibrium solutions corresponding to a Grad-Shafranov equation \citep{Shafranov:1966}. This equation is in general a non-linear partial differential equation and contains three unknown functions, related to the poloidal flux, the poloidal current, and a third one which in the Hall description is related to the advection of the magnetic field by the electron fluid. We develop analytic solutions for Hall equilibria and describe a numerical technique to solve the non-linear version of the Grad-Shafranov equation that self-consistently solves for the boundary of the region of closed poloidal field lines inside the star.

We also discuss the application of our results to magnetostatic equilibria of fluid stars. A Grad-Shafranov type equation can be derived for barotropic fluid stars where gravitational forces, pressure gradients and Lorentz forces are in equilibrium \citep{Prendergast:1956}. Despite the physics being very different, one ends up solving the same variant of Grad-Shafranov equation we find for Hall equilibria.

The long standing question of magnetic equilibria of stars, first discussed in the context of the solar magnetic field \citep{Cowling:1945} and then in more general arguments \citep{Ferraro:1954, Prendergast:1956, Mestel:1956, Roxburgh:1966}, has received much attention recently after  \cite{BraithwaiteSpruit:2004} and \cite{BraithwaiteNordlund:2006} numerically found stable equilibria consisting of mixed poloidal and toroidal fields. Analytical and semi-analytical solutions have been proposed by \cite{Lyutikov:2010} and \cite{Aly:2012}, with \cite{Glampedakis:2012} and \cite{Lander:2012} discussing the importance of the superconducting component of the star. MHD equilibria can also be found through a variation principle \citep{BroderickNarayan:2008, DuezMathis:2010, Duez:2010}. By solving self-consistently for the boundary of the closed field line region, our calculations avoid the issue of the Grad-Shafranov equation being overconstrained as discussed by \cite{Lyutikov:2010}. In agreement with these previous analytic calculations, we find our solutions have a very similar character to the simulation results of \cite{BraithwaiteSpruit:2004} and \cite{BraithwaiteNordlund:2006}. We note however, that the fluid in these simulations is an ideal gas stratified by entropy, so it not barotropic and thus not constrained to satisfy the Grad-Shafranov equation. The similar appearance of the solutions might thus be unrelated to this equation.

An outline of the paper is as follows. The formulation of the Grad-Shafranov equation in the cases of Hall equilibrium and MHD equilibrium is given in \S 2. Our numerical and analytic solutions are presented in \S 3, including a discussion of the rigidly-rotating nature of the electron flow in equilibrium solutions. In \S 4 we discuss the application to magnetic field evolution in neutron stars, including the importance of the differences between MHD and Hall equilibria. We conclude in \S 5. The details of our numerical method are given in the Appendix.

\section{Formulation of the problem}

We first derive the Grad-Shafranov equation for both the Hall and MHD equilibria, and highlight the similarities and differences between the two cases.
We assume an axially symmetric magnetic field configuration. The flux emerging from the surface of the star is the source of a vacuum field  in the exterior that vanishes at infinity. We adopt spherical coordinates $(r, \theta, \phi)$, and $\mu =\cos\theta$. We express the magnetic field in terms of two scalar functions: $\Psi(r, \mu)$, the poloidal magnetic flux, and $cI(r, \mu)/2$, the poloidal current, both of them passing through a spherical cap of radius $r$ and cosine of opening angle $\mu$:
\begin{eqnarray}
\bm{B}=\nabla \Psi \times \nabla \phi + I \nabla \phi\,.
\label{BFIELD}
\end{eqnarray}
Field lines lie on surfaces of constant $\Psi$. By construction, the field is divergence-free. Faraday's induction law is:
\begin{eqnarray}
\frac{\partial \bm{B}}{\partial t}= - c \nabla \times \bm{E}\,.
\label{FARADAY}
\end{eqnarray}
The electric field is $\bm{E}=-\frac{1}{c}\bm{v} \times \bm{B} +\frac{\bm{j}}{\sigma}$, where $\bm{v}$ is the velocity of the electrons, $\bm{j}$ the electric current density and $\sigma$ is the electric conductivity. Given that the electric current is carried by electrons, we write for the current density $\bm{j} =-n_{\rm e}{\rm e} \bm{v}$, where $n_{\rm e}$ is the number density of electrons, and ${\rm e}$ the elementary charge. From Amp\`ere's law, the electric current density is $\bm{j}=\frac{c}{4 \pi} \nabla \times \bm{B}$, and by substitution into equation (\ref{FARADAY}), we find:
\begin{eqnarray}
\frac{\partial \bm{B}}{\partial t} = - \frac{c}{4 \pi {\rm e}}\nabla \times \left( \frac{\nabla \times \bm{B}}{n_{\rm e}} \times \bm{B}\right) -\frac{c^{2}}{4 \pi} \nabla \times \left( \frac{\nabla \times \bm{B}}{\sigma} \right)\,.
\label{BEVOL}
\end{eqnarray}
The first term in the right hand side of equation (\ref{BEVOL}) leads to Hall evolution, where the electric current advects the magnetic field lines, while the second term is the Ohmic dissipation. The typical timescale for Hall evolution is $\tau_{H}\approx  \frac{4 \pi n_{\rm e} {\rm e}L^{2}}{cB}$, while for the competing Ohmic dissipation is $\tau_{Ohm} \approx \frac{4 \pi \sigma L^{2}}{c^{2}}$. The ratio of the two timescales is $\frac{\tau_{Ohm}}{\tau_{H}} \sim 4\times 10^{4} \frac{B_{14}}{T_{8}^{2}}\left(\frac{\rho}{\rho_{\rm nuc}}\right)^{2}$ \citep{GoldreichReisenegger:1992}, where $\rho$ is the mass density, $\rho_{\rm nuc}$ is the nuclear density, $T_{8}$ is the temperature scaled to $10^{8}$K and $B_{14}$ is the magnetic field scaled to $10^{14}$G. Thus there is a range of densities, temperatures and magnetic field intensities where the Hall effect is much faster than Ohmic dissipation. In this range we seek for Hall equilibria.

\subsection{Hall equilibrium}

Assuming infinite conductivity, Hall equilibria are states where the Hall term of equation (\ref{BEVOL}) is equal to zero. Integrating once yields
\begin{eqnarray}
\frac{1}{n_{\rm e}}(\nabla \times \bm{B}) \times \bm{B}=\nabla S\,,
\label{HALL_VECTOR}
\end{eqnarray}
where $S$ is an arbitrary scalar function of $r$ and $\mu$. Substituting the magnetic field from equation (\ref{BFIELD}), the azimuthal component of (\ref{HALL_VECTOR}) is
\begin{eqnarray}
\nabla \Psi \times \nabla I =0 \,,
\end{eqnarray}
which shows that $I=I(\Psi)$. We then obtain the poloidal component of equation (\ref{HALL_VECTOR}):
\begin{eqnarray}
\Big(\frac{\partial^2 \Psi}{\partial r^2} + \frac{1-\mu^{2}}{r^{2}}\frac{\partial^{2} \Psi }{\partial \mu^{2}} + II' \Big)\nabla \Psi \nonumber \\
+r^{2}(1-\mu^{2})n_{\rm e}(r, \mu)\nabla S =0\,,
\label{HALL0}
\end{eqnarray}
where a prime denotes differentiation with respect to $\Psi$. From equation (\ref{HALL0}) we deduce that $\nabla \Psi \parallel \nabla S$, so $S=S(\Psi)$. Defining the Grad-Shafranov operator $\Delta^{*} =\frac{\partial^{2}}{\partial r^{2}} +\frac{1-\mu^{2}}{r^{2}}\frac{\partial^{2}}{\partial \mu^{2}}$, the Hall equilibrium equation (\ref{HALL0}) reduces to
\begin{eqnarray}
\Delta^{*}\Psi +II' +r^{2}(1-\mu^{2})n_{\rm e}(r, \mu)S'=0\,.
\label{HALL}
\end{eqnarray}
In the mathematical derivation, the electron number density $n_{\rm e}$ can in principle be a function of $r$ and $\mu$ and is related to the mass density $\rho$ of the star, so that $n_{\rm e}=\rho Y_{\rm e}$, where $Y_{\rm e}$ is the electron number per unit mass. For realistic cases the dependence on the radius will be much stronger than the angular dependence, so we shall use $n_{\rm e}=n_{\rm e}(r)$ throughout this paper. Note that we have not taken into account relativistic terms, as $v/c \ll 1$. We have also assumed that conductivity is large enough so that Ohmic dissipation can be neglected in first approximation; its subdominant effect is discussed in \S 4.2.

\subsection{Barotropic MHD equlibrium}

In barotropic matter, pressure $P$ is a function of the density $P=P(\rho)$ only. Force equilibrium is given by
\begin{eqnarray}
\frac{1}{c \rho} \bm{j} \times \bm{B} = \frac{1}{\rho}\nabla P + \nabla \Phi\,,
\label{BAR1}
\end{eqnarray}
where $\Phi$ is the gravitational potential. Taking the curl of the above equation, using the fact that $\nabla P$ is parallel to $\nabla \rho$, and substituting $\bm{B}$ for $\bm{j}$ from Amp\`ere's law, we find
\begin{eqnarray}
\nabla \times \Big( \frac{\nabla \times \bm{B}}{\rho}  \times \bm{B} \Big)=0\,.
\label{BAR1_5}
\end{eqnarray}
Integrating once and with $S_{B}$ an arbitrary scalar
\begin{eqnarray}
\frac{1}{\rho} (\nabla \times \bm{B})\times \bm{B}=\nabla S_{B}\,,
\label{BAR2}
\end{eqnarray}
which is mathematically identical to the equation we have found for the Hall equilibrium (\ref{HALL_VECTOR}). Following the same line of arguments as in the Hall case, it reduces to
\begin{eqnarray}
\Delta^{*}\Psi +II' +r^{2}(1-\mu^{2}) \rho(r, \mu)S_{B}'=0\,.
\label{BARTROP}
\end{eqnarray}
The physical interpretation, however, is very different, as for instance here the mass density $\rho$ plays the role played by the electron density $n_{\rm e}$ is Hall case, and $S$ and $S_{B}$ have different physical meaning.  
Although the equations governing the equilibrium in both cases are nearly identical, the equations for the time-evolution of small perturbations are completely different, and therefore stability or instability of an equilibrium in one case can by no means be extrapolated to the other. We discuss this key difference further in \S 4.

\section{Solutions}

In this section, we calculate equilibrium solutions for the magnetic field. After discussing analytical solutions in \S 3.1, we present a new numerical method for calculating equilibria in \S 3.2. In \S 3.3, we discuss the current distribution and electron velocity profile, which is particularly important for Hall evolution with magnetic field lines frozen into the electron fluid.

Before attempting to solve for equilibria we first outline some general properties of magnetic fields, which constrain our solutions. Both components of the field normal and parallel to the boundary must be continuous to ensure that the field is everywhere divergence-free and there are no current sheets. Even if current sheets had formed at some stage, we expect they had dissipated fast enough and they are not present in a steady-state solution.

\begin{figure}
\centering
\includegraphics[width=0.50\columnwidth]{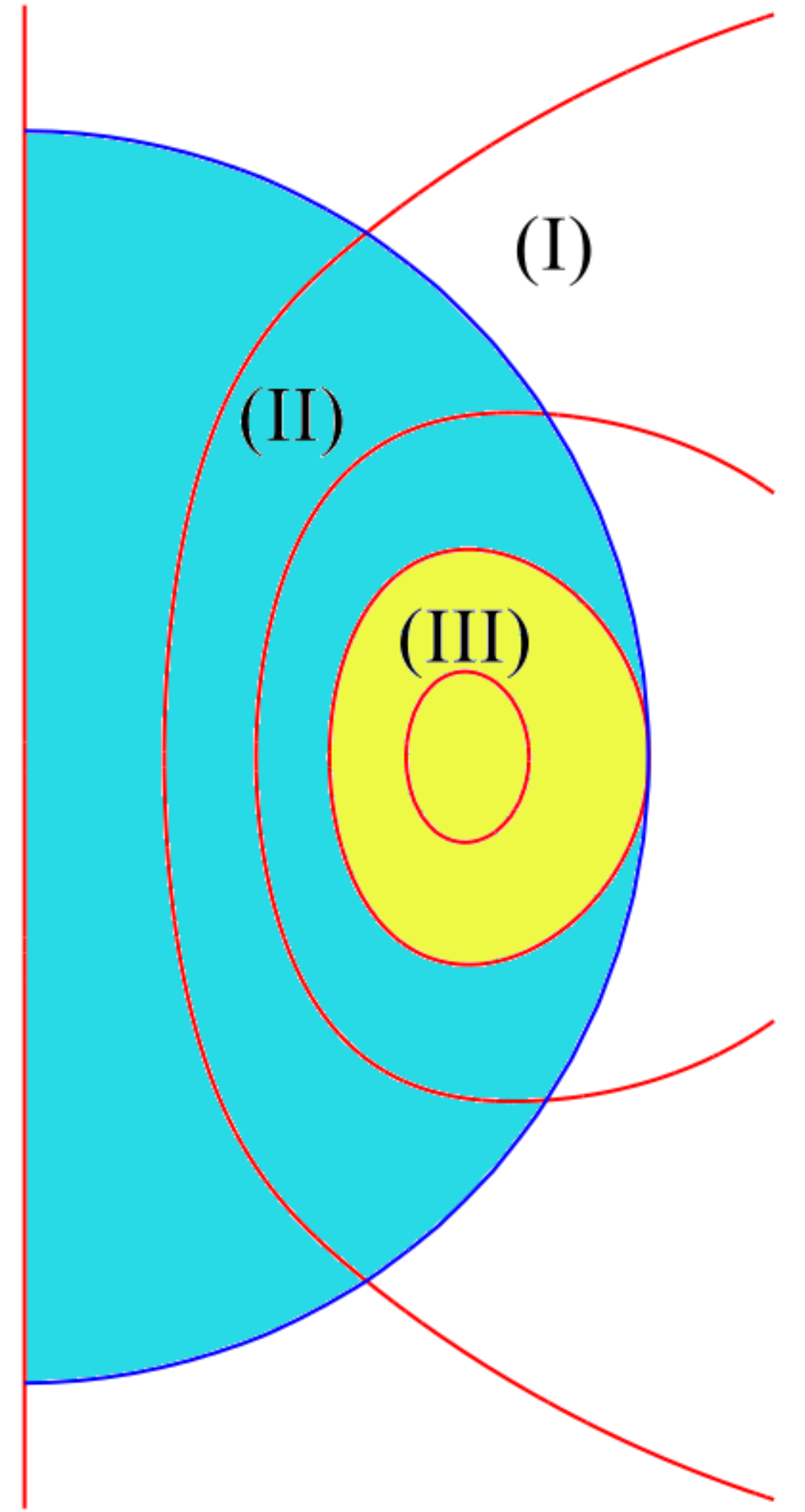}
\caption{Meridional cut of the star showing schematically its poloidal field lines and the different regions defined by them. Region I is the vacuum outside of the star where the field is current-free. Region II is the part of the star containing field lines that are connected to region I; in this region, there cannot be any toroidal field, but it is not current-free. Region III is the part of the star containing field lines that are not connected to Region I and close inside the star; in this region it is possible to have both toroidal and poloidal fields and currents. Note that the field in this picture reaches the centre of the star; as it would be permitted for (barotropic) MHD equilibria. Hall equilibria, however, need to be excluded from the fluid core of the star to be physically meaningful.}
\label{Fig:1}
\end{figure}

We separate the domain where we solve the equation into three regions, as shown in Figure \ref{Fig:1}. Region I is the exterior of the star, where the field is current-free and consequently there cannot be any toroidal field. Region II is the part of the interior of the star containing field lines that penetrate the surface and connect to the field of region I. These field lines cannot accommodate a toroidal field either, because $I=I(\Psi)$ implies $I$ is constant along field lines. 
Region III contains the field lines inside the star that are not connected to the external field and form closed tori; these field lines may have both toroidal fields and toroidal currents. In region I, we have $I=0$ and $n_{\rm e}=0$, thus the term proportional to $S'$ is eliminated. In region II, $I=0$,  and in region III there is no constraint for $S$ or $I$, and they are in general non-zero. Therefore in the three regions, $\Psi$ satisfies:
\begin{eqnarray}
\begin{array}{l l}
\Delta^{*}\Psi=0, & \,{\rm I}\,, \\
\Delta^{*}\Psi+r^{2}(1-\mu^2)n_{\rm e}(r)S'=0, & \, {\rm II}\,, \\
\Delta^{*}\Psi+II'+r^{2}(1-\mu^2)n_{\rm e}(r)S'=0, & \, {\rm III}\,,
\end{array}
\label{GS}
\end{eqnarray} 
respectively.

\subsection{Analytical solutions}

We seek analytical solutions with the method of separation of variables. We assume that the flux function is a product of a radial and an angular function. Then we choose the functional dependence of $I(\Psi)$ and $S(\Psi)$ so that they allow such separable solutions. Finally, we match the solutions of the various regions to ensure continuity along the boundaries. We shall present two families of analytical solutions that satisfy the requirements we have set: solutions with purely poloidal fields (\S \ref{POLOIDAL}); and solutions with both poloidal and toroidal fields fully confined inside the star (\S \ref{CONFINED}). We have not been able to solve for a more general configuration analytically, thus we explore numerical solutions in \S \ref{NUMERICAL}. 

\subsubsection{Purely poloidal fields}
\label{POLOIDAL}

In this solution we have no toroidal field, thus $I=0$ and the magnetic field inside the star satisfies equation (\ref{GS}-II) everywhere inside the star. The external vacuum satisfies equation (\ref{GS}-I).

\begin{figure}
\centering
a\includegraphics[width=0.47\columnwidth]{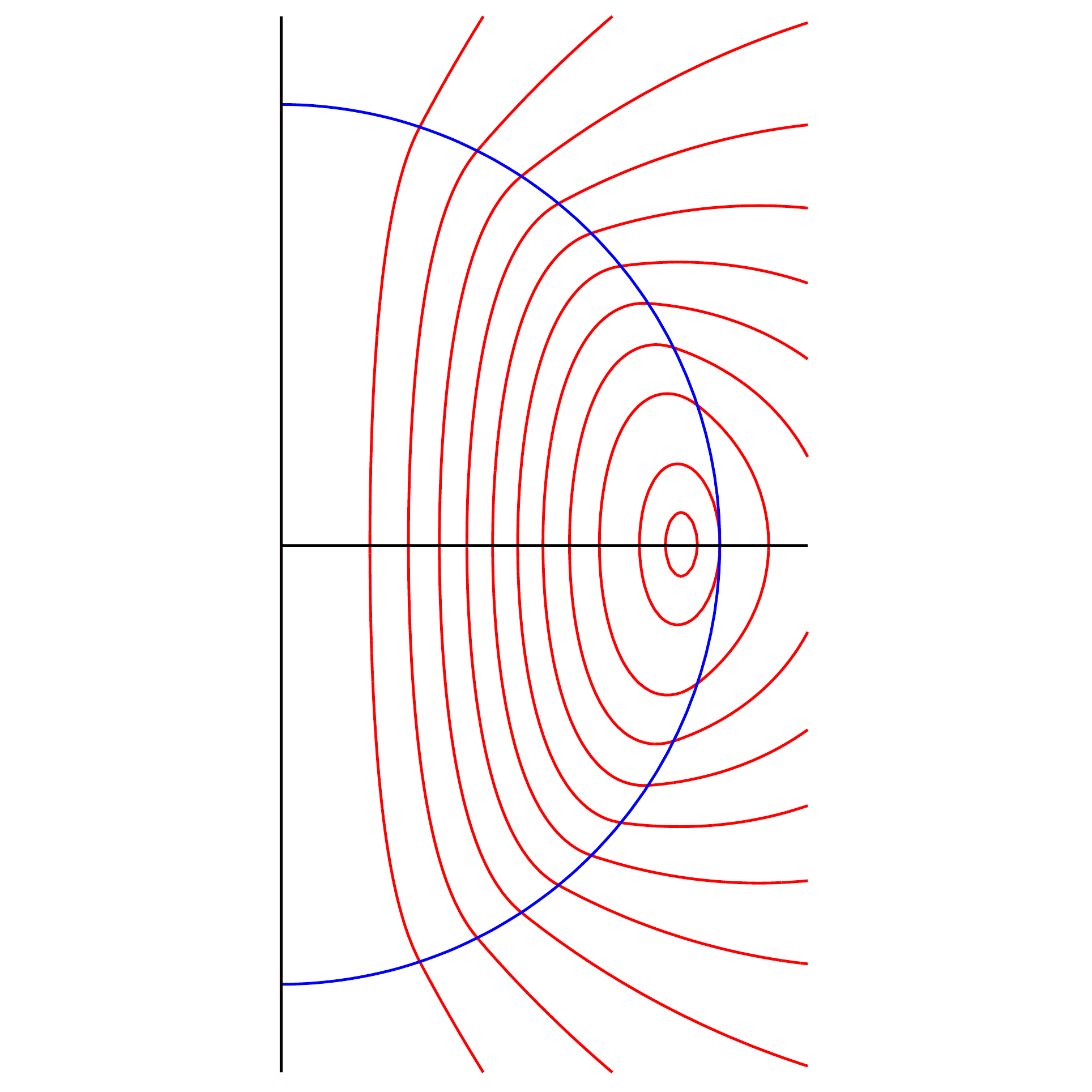}
b\includegraphics[width=0.47\columnwidth]{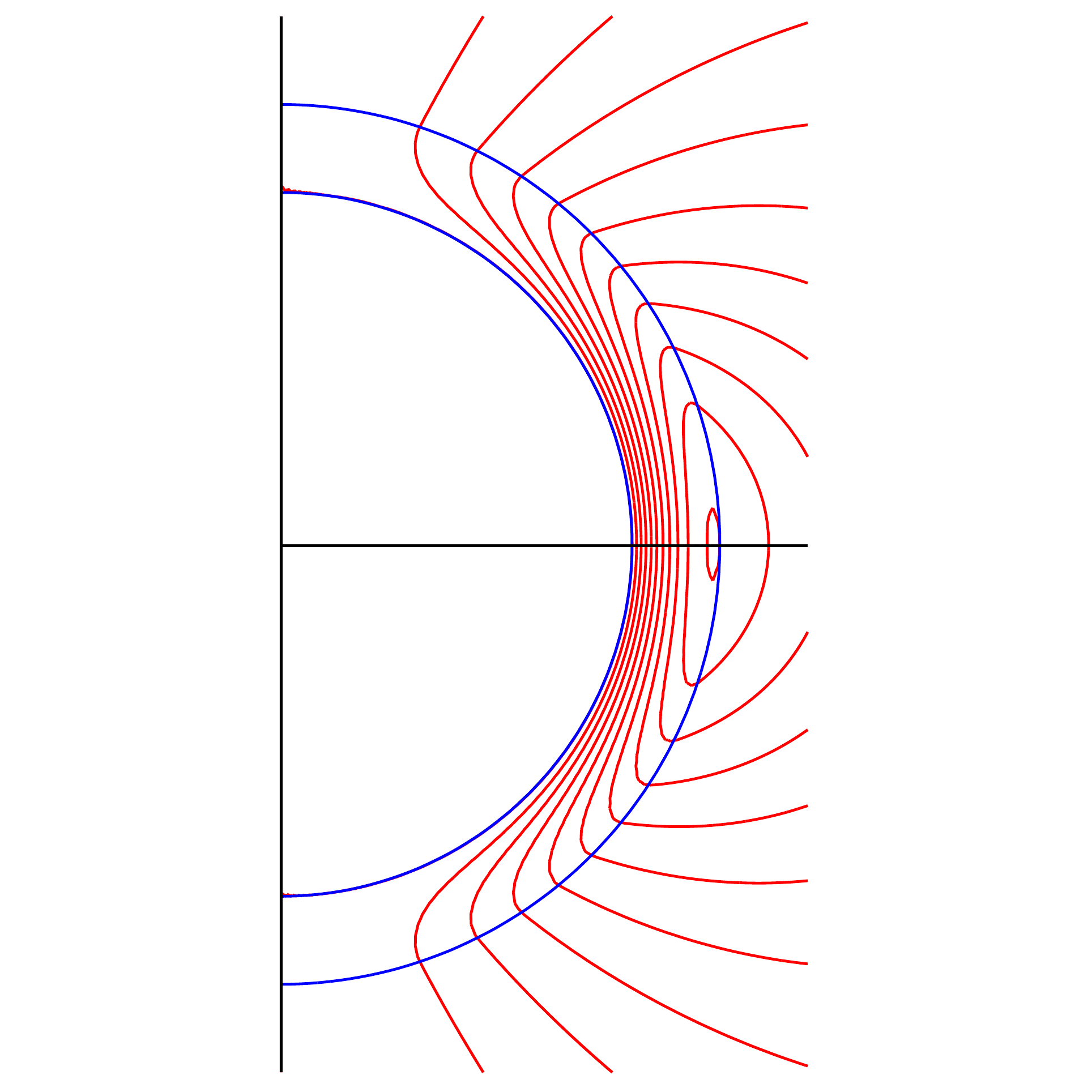}
c\includegraphics[width=0.47\columnwidth]{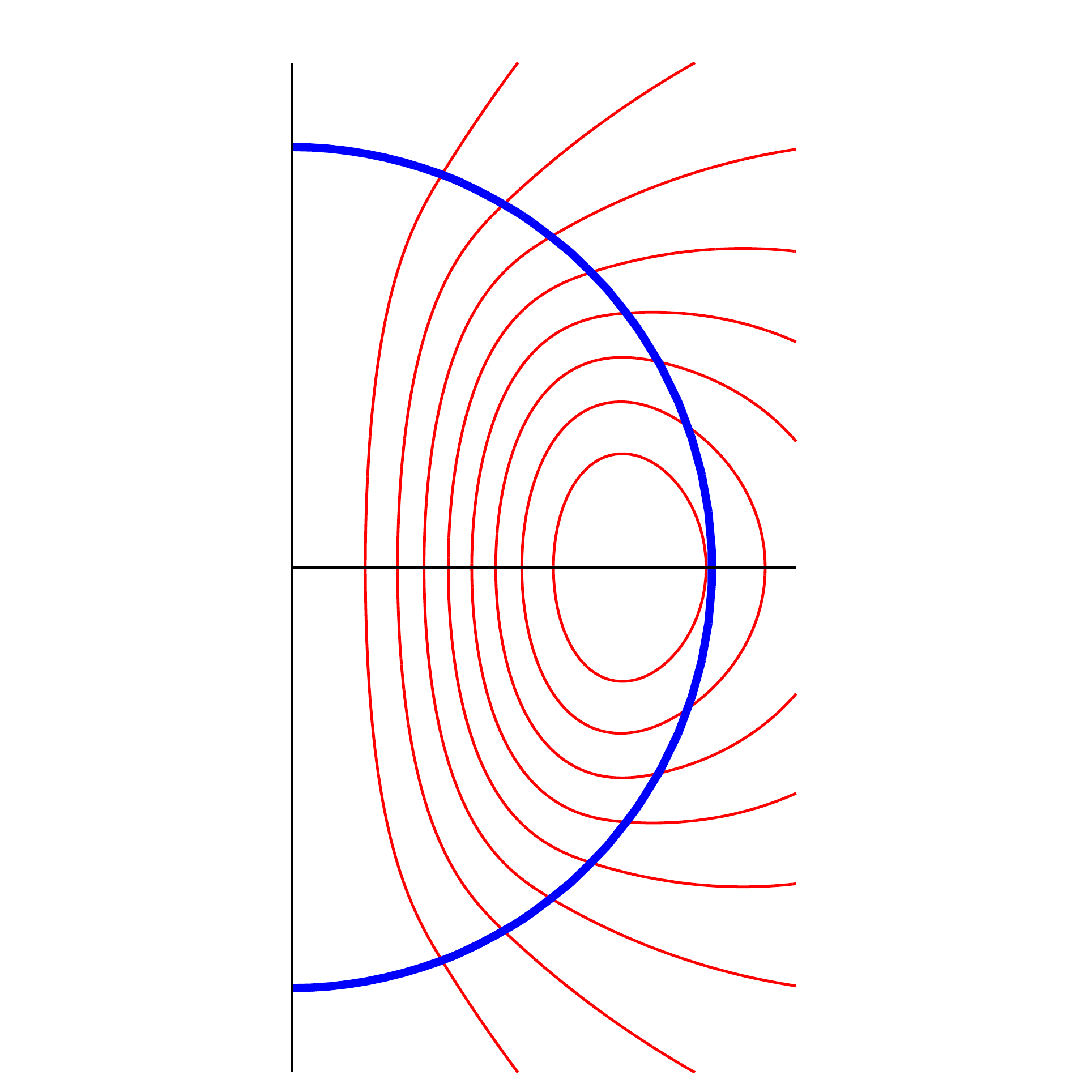}
d\includegraphics[width=0.47\columnwidth]{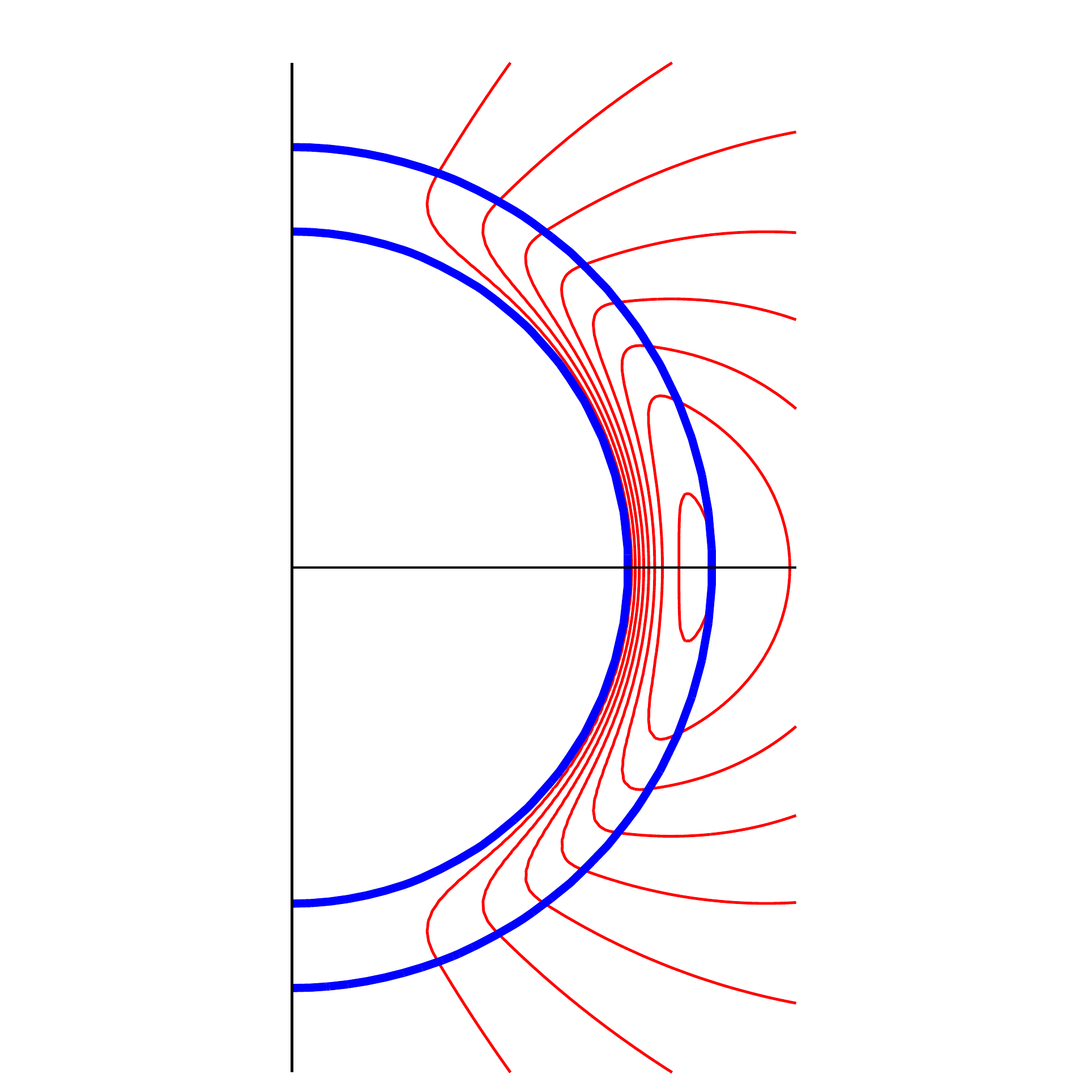}
\caption{The structure of the field for the analytical solutions of equation (\ref{PsiPol}). In all examples the field is purely poloidal. In cases (a) and (b) the electron number density is uniform while in (c) and (d) it varies as  $n_{\rm e}\propto (r_{*}^2-r^{2}) $. In (a) and (c) the field is allowed to reach the centre of the star, while in (b) and (d) it is excluded from the core starting at $r_{in}=0.8r_{*}$.}
\label{Fig:2}
\end{figure}

We solve equation (\ref{GS}) in region I by separation of variables in $r$ and $\mu$, writing for the solution $R_{\rm I}(r)M_{\rm I}(\mu)$. We find $M_{{\rm I},\ell}=(1-\mu^{2})\frac{d P_{\ell}(\mu)}{d \mu}$, $P_{\ell}$ being the $\ell$th-order Legendre polynomial and $R_{{\rm I},\ell}=c_{I,\ell}r^{-\ell}+d_{I,\ell}r^{\ell+1}$, with $d_{I,\ell}=0$ to avoid divergence at infinity. The solution is a sum of the products of $R_{{\rm I},\ell}M_{{\rm I},\ell}$. Non-trivial separable solutions inside the star are possible for $S^{\prime}=S_{\rm II}$, where $S_{\rm II}$ is a constant. This choice provides a useful example, as it can be well understood and used to test numerical solutions. We solve the following equation:
\begin{eqnarray}
\Delta^{*}\Psi + r^{2} (1- \mu^{2}) n_{\rm e}(r) S_{\rm II}=0\,.
\end{eqnarray}
There is one term that does not depend on $\Psi$, so the equation is inhomogeneous. Its general solution is a sum of the general solution of its homogeneous counterpart (obtained by eliminating the term in question) plus any particular solution of the full, inhomogeneous equation. Matching the fields from regions I and II without discontinuities, and thus no surface currents, impose the conditions $\frac{\partial \Psi}{\partial r}|_{r_{*}^{+}}=\frac{\partial \Psi}{\partial r}|_{r_{*}^{-}}$ and $ \Psi |_{r_{*}^{+}}= \Psi |_{r_{*}^{-}}$. The flux has to be zero at the inner boundary thus $\Psi(r_{in})=0$, where $r_{in}$  is the deepest point the field threads the star. For a star where the field reaches the centre it is $r_{in}=0$, but it can be much larger for a star where the field is confined in a thin crust. The resulting solution is
\begin{eqnarray}
\begin{array}{l l}
\Psi=c_{I,1}\frac{(1-\mu^{2})}{r}, & {\rm I}\,,  \\
\Psi=(1-\mu^{2})\Big(c_{II,1} r^{2}+ d_{II,1}r^{-1} +\\
\frac{S_{\rm II}}{3r} \Big[ \int_{r_{in}}^{r} n_{\rm e}(\tilde{r}) \tilde{r}^{4} d\tilde{r} -r^{3}\int_{r_{in}}^{r} n_{\rm e}(\tilde{r})\tilde{r}d\tilde{r}\Big] \Big) & {\rm II,~III}\,.
\end{array}
\label{PsiPol}
\end{eqnarray}
In this solution there are four constants to be determined $(c_{I,1},~c_{II,1},~d_{II,1},~S_{\rm II})$. They are found using the continuity and boundary conditions outlined above and the freedom of the overall normalization of the magnetic flux emerging from the star. The results are plotted in Figure \ref{Fig:2}. In the assumption of constant $n_{\rm e}$, $r_{in}=0$ and $r_{*}=1$ we find that it is $c_{I,1}=\frac{1}{15}S_{\rm II}n_{\rm e}$,  $c_{II,1}=\frac{1}{6} S_{\rm II}n_{\rm e}$ and $d_{II,1}=0$.

In the cases where fields are confined in a thin crust and do not reach the centre of the star, the inner boundary of the crust has a current sheet as the field is pushed out of the arguably super conducting core; such fields will have a non-zero $\frac{\partial \Psi}{\partial r}|_{r_{in}}$. Another important consequence is that the magnetic field in the crust is stronger than the inferred dipole field by about an order of magnitude  $B_{\theta} \sim r_{*}/(r_{*}-r_{in}) B_{r}$, because the field lines are confined within the crust, providing a large reservoir of accessible magnetic energy.

\subsubsection{Fully confined fields}
\label{CONFINED}

Analytical solutions with both poloidal and toroidal field are possible if the field is fully confined within the star, thus region III covers all the star. In this solution there is no field in region I.

The field in region III has a non-zero $I(\Psi)$, in addition to the non-zero $S(\Psi)$. Analytical separable solutions can be found for $I=\alpha (\Psi-\Psi_{0})$, and $S=S_{0}+S_{\rm III}(\Psi-\Psi_{0})$; where $\Psi_{0}$ is the value of $\Psi$ on the boundary of region III. Under these assumptions the solution for equation (\ref{GS}-III) is
\begin{eqnarray}
\Psi=\sum\limits_{\ell=1}^{\ell=\infty} \Big\{r^{1/2}[g_{1,\ell}J_{\ell +1/2}(\alpha r) + g_{2,\ell}Y_{\ell +1/2}(\alpha r)] \nonumber \\
\times (1-\mu^{2})P_{\ell}'(\mu) \Big\}\nonumber \\
-\frac{\pi S_{\rm III} r^{1/2}(1-\mu^{2})}{2}\Big[Y_{3/2}(\alpha r) \int_{r_{in}}^{r} \tilde{r}^{5/2}J_{3/2}(\alpha \tilde{r}) n_{\rm e}(\tilde{r}) d\tilde{r} \nonumber \\
-J_{3/2}(\alpha r) \int_{r_{in}}^{r} \tilde{r}^{5/2}Y_{3/2}(\alpha \tilde{r}) n_{\rm e}(\tilde{r}) d\tilde{r}\Big] +\Psi_{0}\,,
\label{RIII}
\end{eqnarray}
where $J$ and $Y$ are Bessel functions \citep{Abramowitz:1972}. In this particular solution it is $\Psi_{0}=0$.

\begin{figure}
\centering
a\includegraphics[width=0.47\columnwidth]{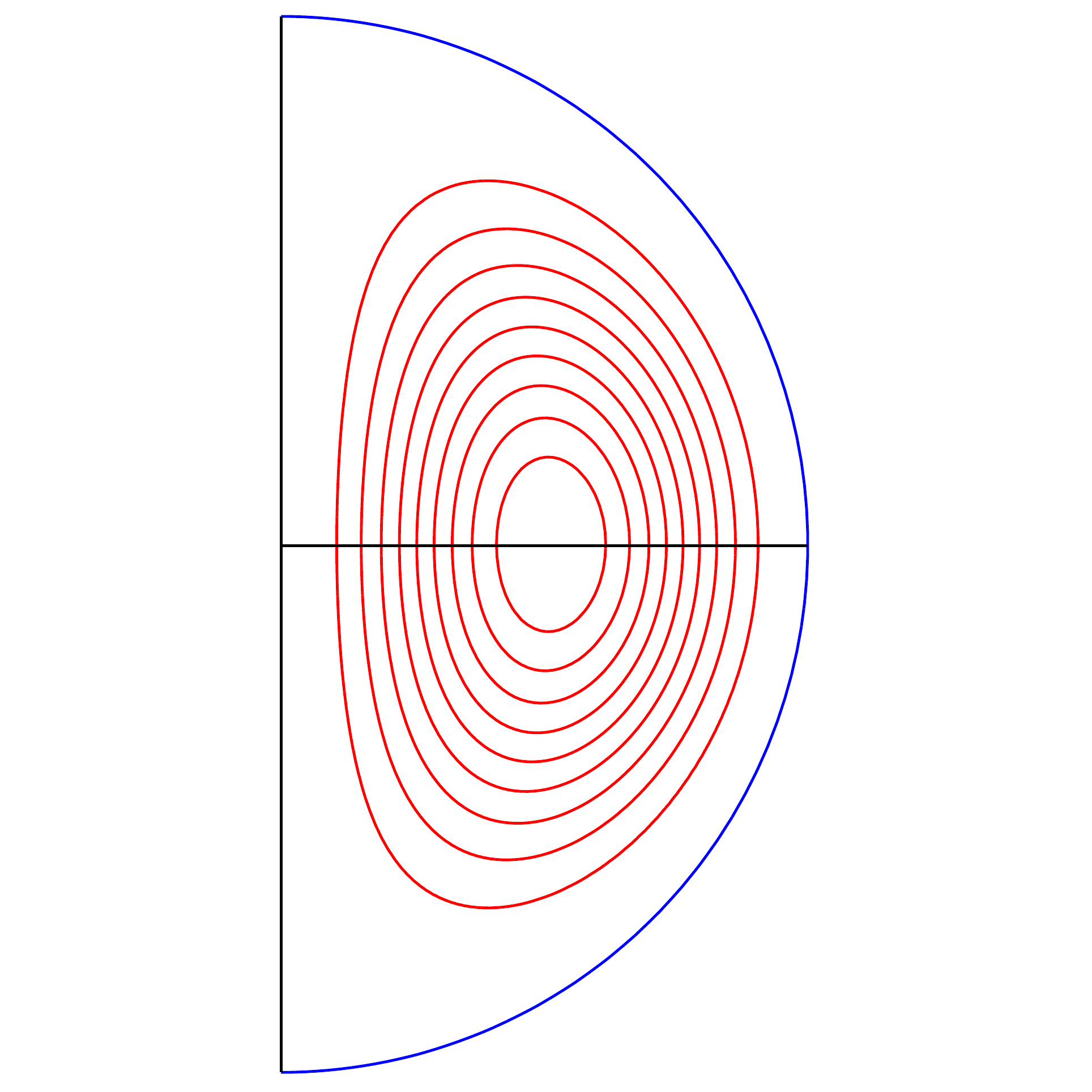}
b\includegraphics[width=0.47\columnwidth]{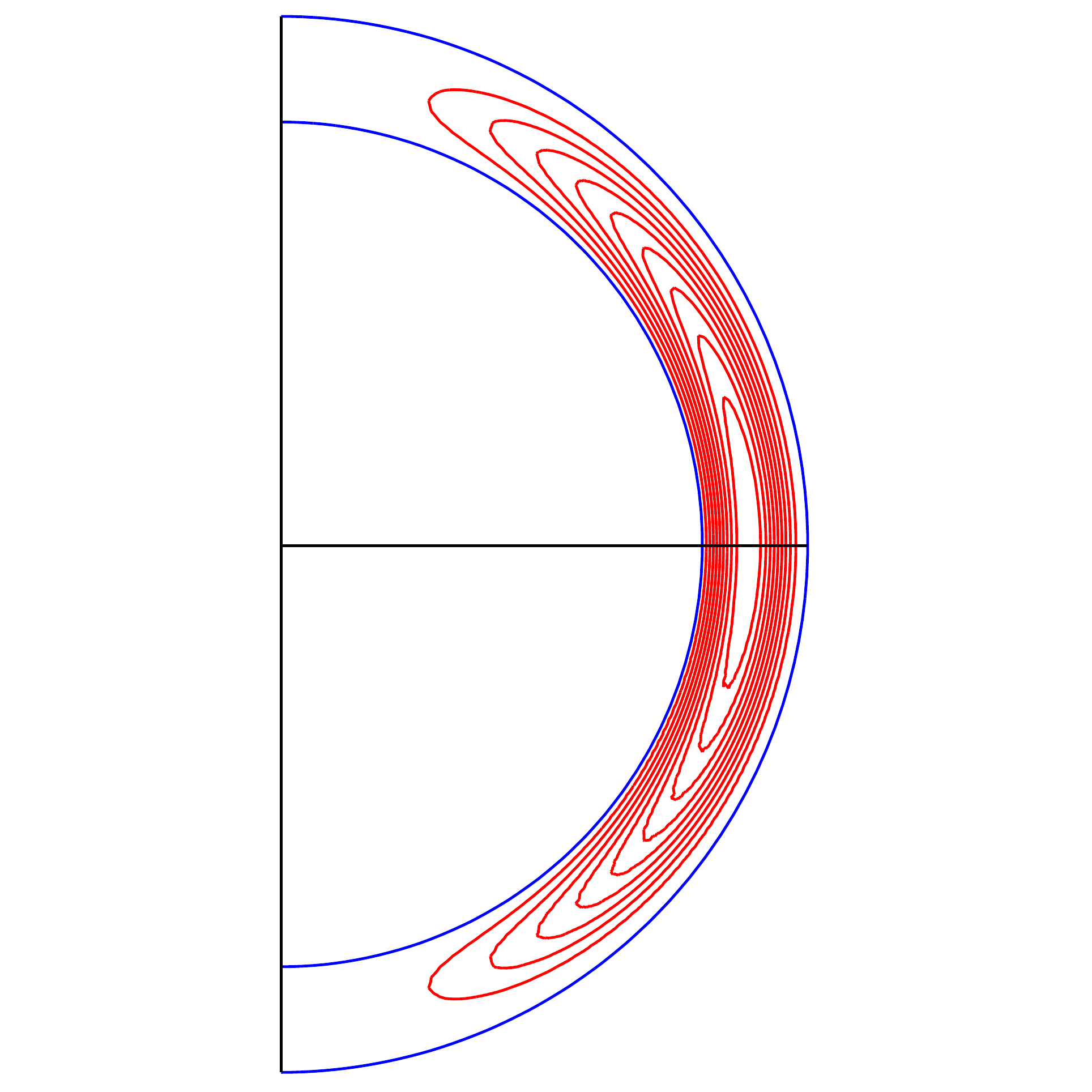}
\caption{The confined solution to equation (\ref{RIII}) as described in Section \ref{CONFINED} for uniform electron density, which contains both poloidal and toroidal fields confined in the star. Poloidal field lines, which are also sections of surfaces of constant poloidal flux $\Psi$ and constant $I$, are plotted. The field reaches the centre of the star in case (a) and  is confined in a crust whose inner radius is $0.8r_{*}$ in (b).}
\label{Fig:3}
\end{figure}

Requiring the field to vanish on the surface of the star with no surface currents, thus $\Psi(r_{*})=0$ and $\frac{\partial \Psi}{\partial r}|_{r=r_{*}}=0$, leads us to keep only the $\ell=1$ term from the sum. This is because the inhomogeneous term is proportional to $(1-\mu^{2})$ and can be arranged so that it can cancel the contribution of the homogeneous dipole term on the surface, while the non-dipole terms, because of orthogonality, cannot. There are four constants left $(\alpha,~ g_{1,1}, ~g_{2,1},~ S_{\rm III})$, that can be determined using the above boundary conditions, the inner boundary condition $\Psi(r_{in})=0$  and the overall normalization of the flux contained inside the star. Note that if we had required $\frac{\partial \Psi}{\partial r}|_{r=r_{in}}=0$ the problem would have be overdetermined and insoluble, unless $r_{in}=0$ which leads to solutions which are proportional to $r^{2}$ near $r=0$. For the case of constant density $n_{e}$ and $r_{in}=0$ the solution is
\begin{eqnarray}
\Psi= \Big[\Big(\frac{\sqrt{2}g_{1}}{\sqrt{\pi \alpha}} +\frac{3S_{\rm III}n_{\rm e}}{\alpha^{4}}\Big)\Big(\frac{\sin(\alpha r)}{\alpha r} -\cos(\alpha r)\Big)\nonumber \\
-\frac{S_{\rm III}n_{\rm e}r^2}{\alpha^2}\Big](1-\mu^2)\,.
\end{eqnarray}
Applying the boundary conditions at $r_{*}=1$, we find that $\tan\alpha =3\alpha/(3-\alpha^{2})$ and $g_{1,1} = [\pi/(2\alpha^{7})]^{1/2} \left( \frac{\alpha^{3} +3\alpha \cos\alpha -3\sin\alpha}{\sin\alpha -\alpha\cos\alpha} \right) S_{\rm III}n_{\rm e}$. This gives an infinite sequence of solutions with progressively more nodes in the domain. The first of these, with no internal nodes, is $\alpha= 5.763$ and $g_{1,1}=-0.1031S_{\rm III}n_{\rm e} $. This solution is plotted in Figure \ref{Fig:3}.a along with a solution where the field is only confined in the crust, Figure \ref{Fig:3}.b. In this configuration, the toroidal field contains $58.3\%$ of the total magnetic energy in the case of the full sphere and $70.6\%$ for the case of the crust. This solution has been discussed in the context of stellar fields \citep{Roxburgh:1966, DuezMathis:2010} and also in the context of magnetic bubbles \citep{Gourgouliatos:2010, Gourgouliatos:2012}.

\subsection{Numerical Solutions}
\label{NUMERICAL}

We have developed a relaxation scheme implementing the Gauss-Seidel method for elliptic differential equations \citep{Press:1992}. The details of the numerical scheme appear in the Appendix. We solve equation (\ref{GS}) everywhere in the star and its surroundings we choose
\begin{eqnarray} 
I=\alpha (\Psi-\Psi_{0})^{\zeta}\,, \Psi>\Psi_{0}\, \nonumber \\ 
I=0\,, \Psi\leq \Psi_{0}\,.
\end{eqnarray}
where $\Psi_{0}=\Psi(r_{*}, \pi/2)$, the value of the flux function on the equator of the star, which is determined self-consistently in each iteration, thus determining the boundary of regions II and III as the locus $\Psi(r, \theta)= \Psi_{0}$. A choice of $\zeta=1.1$ allows the comparison with \cite{Lander:2009}. We have also solved for $I\propto \Psi(\Psi-\Psi_{0})$ and we find that the results are qualitatively similar. We choose a linear form for $S=S_{0}+S_{\rm II}(\Psi-\Psi_{0})$ in region II and $S=S_{0}+S_{\rm III}(\Psi-\Psi_{0})$ in region III. 

In the examples we present, we have chosen $n_{\rm e}=$const. and $n_{\rm e}\propto (r_{*}^2-r^{2})$, the \cite{Tolman:1939} solution of Einstein field equations for spheres of fluid, which is a good approximation for many equations of state of neutron stars \citep{Lattimer:2001} and is close to the mass density profile for an $n=1$ polytrope $P=\rho^{2}$. The latter form of $n_{\rm e}(r)$ shares the important property that  the crust electron density drops by orders of magnitude as one approaches the outer edge of the crust. We have explored the parameter space by choosing various combinations of $\alpha$, $S_{\rm II}$ and $S_{\rm III}$. We have performed our solutions in both spherical and cylindrical coordinates, going up to resolution $400 \times 400$ and finding consistent results. 

\subsubsection{Reproduction of analytical solutions}

\begin{figure}
\centering
\includegraphics[width=0.94\columnwidth]{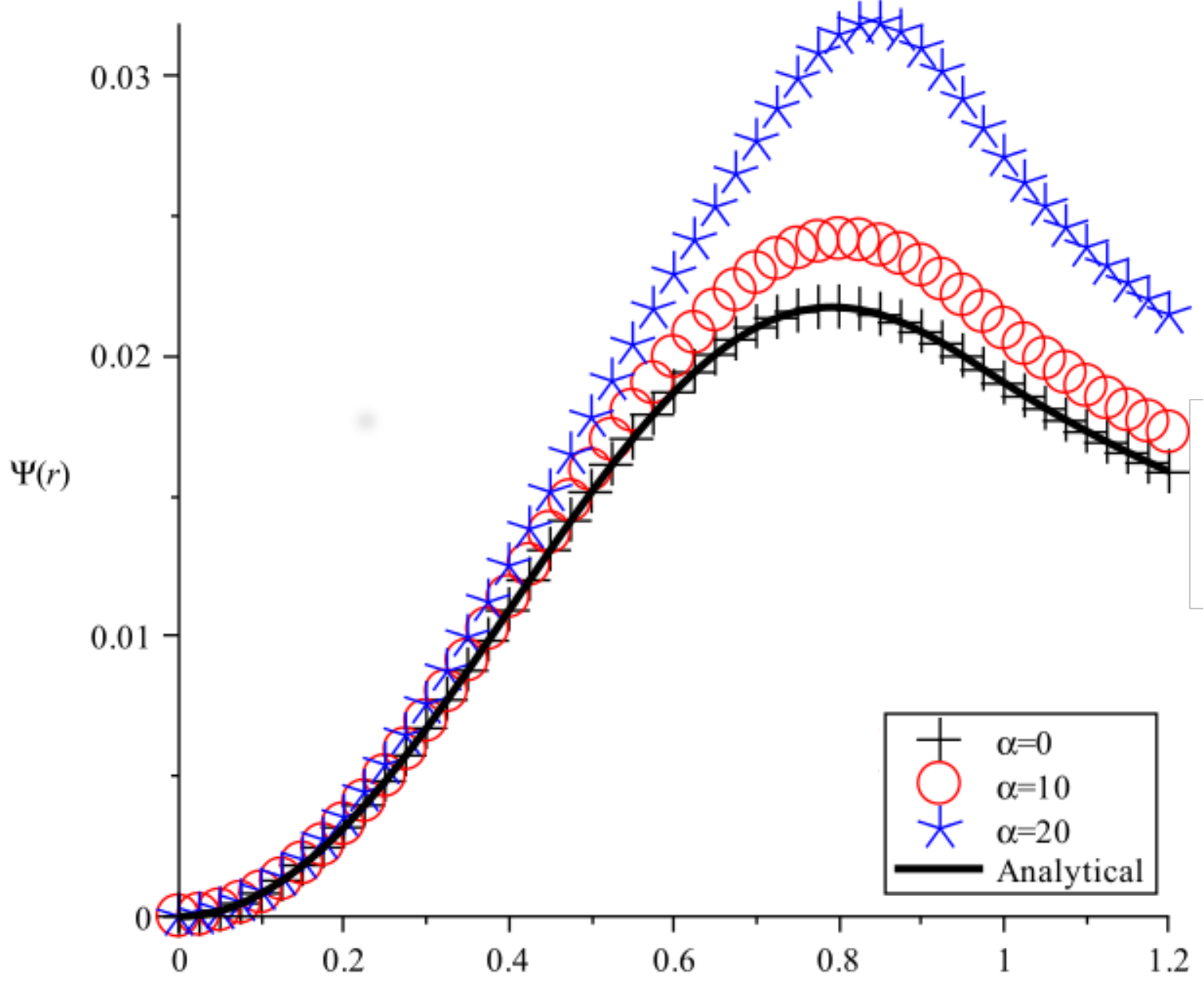}
\caption{Numerical and analytical solutions of $\Psi(r)$ at the equator $\mu=0$ for $I=\alpha(\Psi-\Psi_{0})^{1.1}$,  $S_{\rm II}=S_{\rm III}=1$, $n_{\rm e}=(1-r^{2})$, $r_{in}=0$ and $r_{*}=1$ \. The analytical solution of a purely poloidal field is the solid line, while the black crosses are the numerical solution for the same case ($\alpha=0$). The red circles are the numerical solution corresponding to $\alpha=10$. The blue asterisks are the numerical solution for $\alpha=20$. Larger values of $\alpha$, corresponding to stronger toroidal fields, steepen the profile.}
\label{Fig:4}
\end{figure}

\begin{figure}
\centering
\includegraphics[width=0.94\columnwidth]{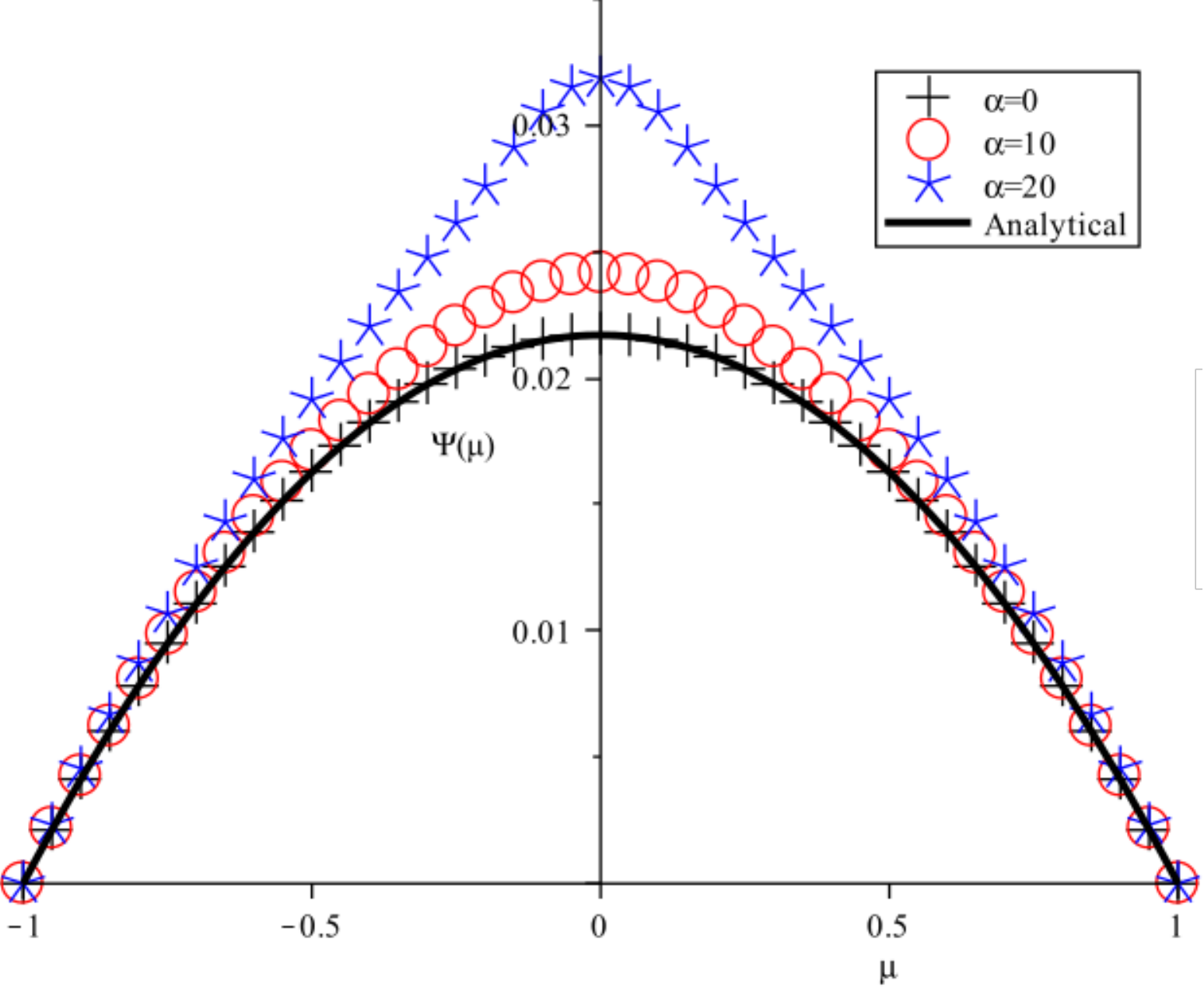}
\caption{Angular dependence of the solutions presented in Figure \ref{Fig:4}. The value of $\Psi(r_{\rm max}, \mu)$ is plotted, where $r_{\rm max}$ is the radius at which the global maximum of $\Psi$ occurs. The symbols are the same as in Figure \ref{Fig:4}. Larger values of $\alpha$, corresponding to stronger toroidal fields, steepen the profile.} 
\label{Fig:5}
\end{figure}

\begin{figure}
\centering
\includegraphics[width=0.94\columnwidth]{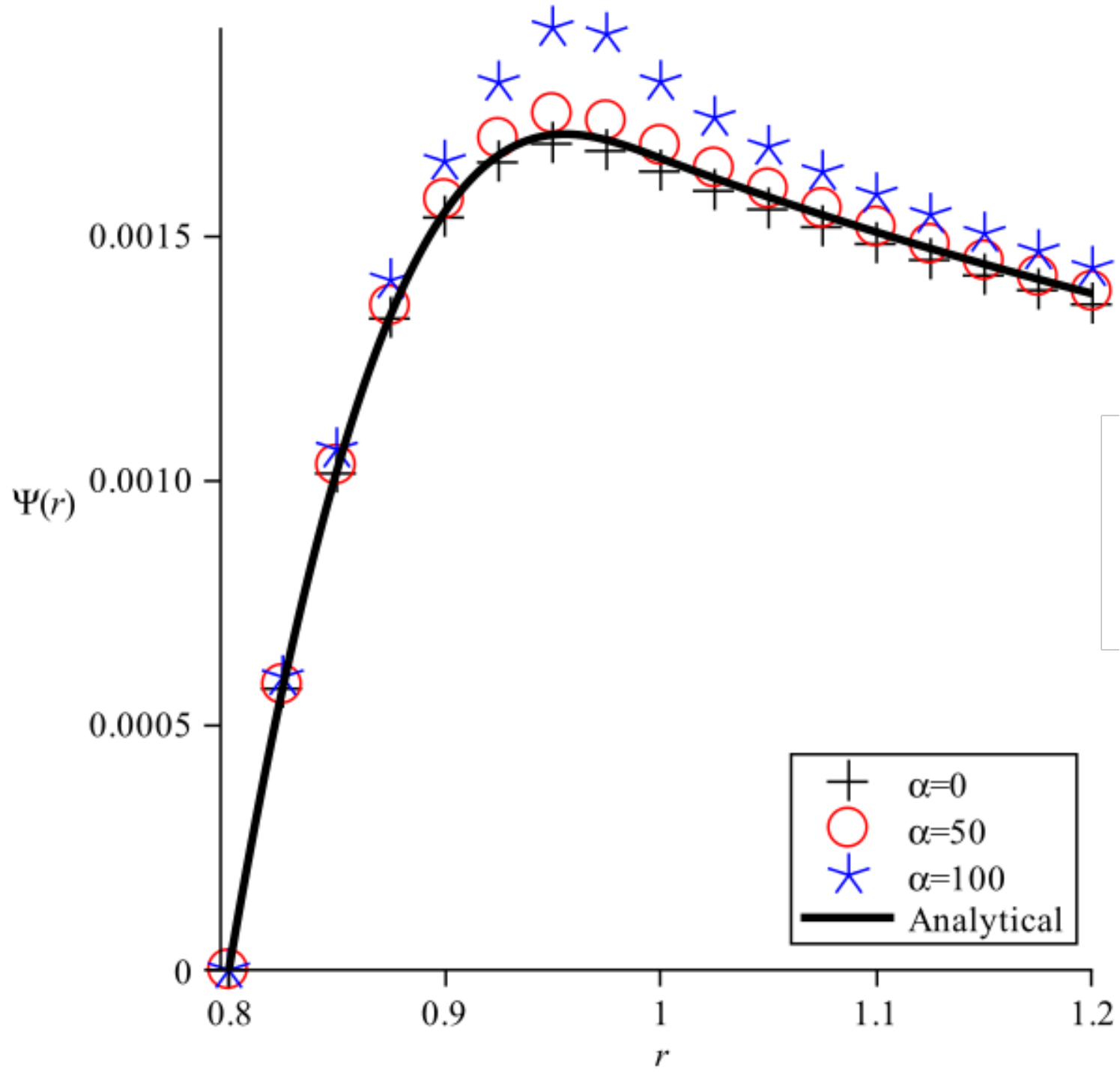}
\caption{Numerical and analytical solutions of $\Psi(r)$ at the equator $\mu=0$ for crust confined fields, for $I=\alpha(\Psi-\Psi_{0})^{1.1}$, $S_{\rm II}=S_{\rm III}=1$, $n_{\rm e}=(1- r^{2})$, $r_{in}=0.8$ and $r_{*}=1$. The solid line is the analytical solution of the purely poloidal field, while the black crosses show the solution for the same case. The red circles is the numerical solution corresponding to $\alpha=50$. The blue asterisks are the numerical solution for $\alpha=100$. Similarly to the solution for the full star, larger values of $\alpha$, corresponding to stronger toroidal fields, lead to steeper profile. }
\label{Fig:6}
\end{figure}

\begin{figure}
\centering
\includegraphics[width=0.94\columnwidth]{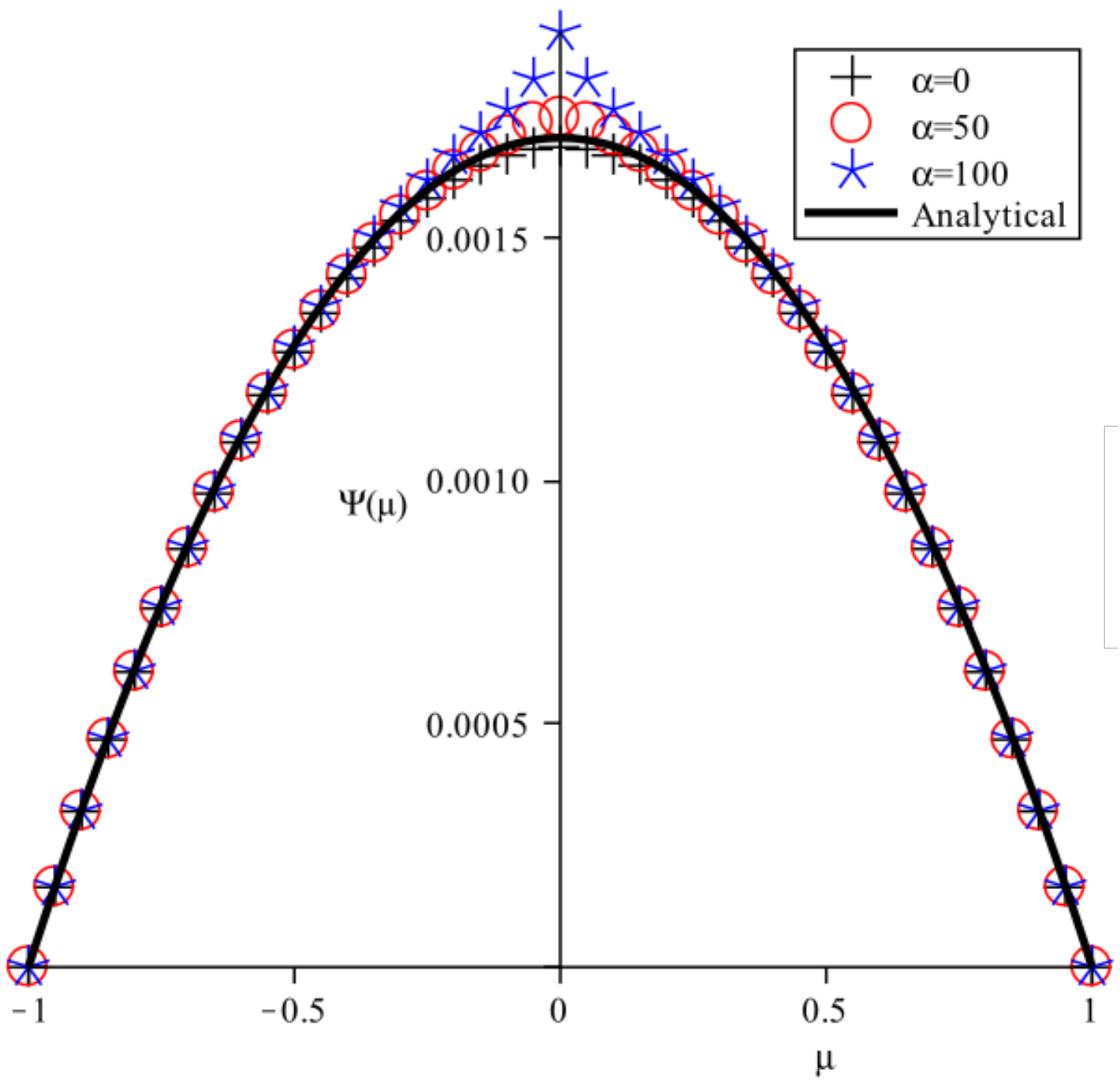}
\caption{Angular dependence of the solutions presented in Figure \ref{Fig:6}. The value of $\Psi(\mu, r_{\rm max})$ is plotted, where $r_{\rm max}$ is the radius at which the maximum of $\Psi$ occurs. The symbols are the same as in Figure \ref{Fig:6}. Similarly to the solution for the full star, larger values of $\alpha$, corresponding to stronger toroidal fields, lead to steeper profile. }
\label{Fig:7}
\end{figure}

To test the accuracy of our numerical scheme, we have solved for the field corresponding to the dipole without toroidal field given by equation (\ref{PsiPol}). We find that the numerical scheme converges to the analytical solution, see Figures \ref{Fig:4} -- \ref{Fig:7}, where the solid lines are the analytical solution and the crosses show the numerical result for the same parameters.

\subsubsection{Solutions with toroidal field}
\label{SOL_NUM}

\begin{figure}
\centering
a\includegraphics[width=0.45\columnwidth]{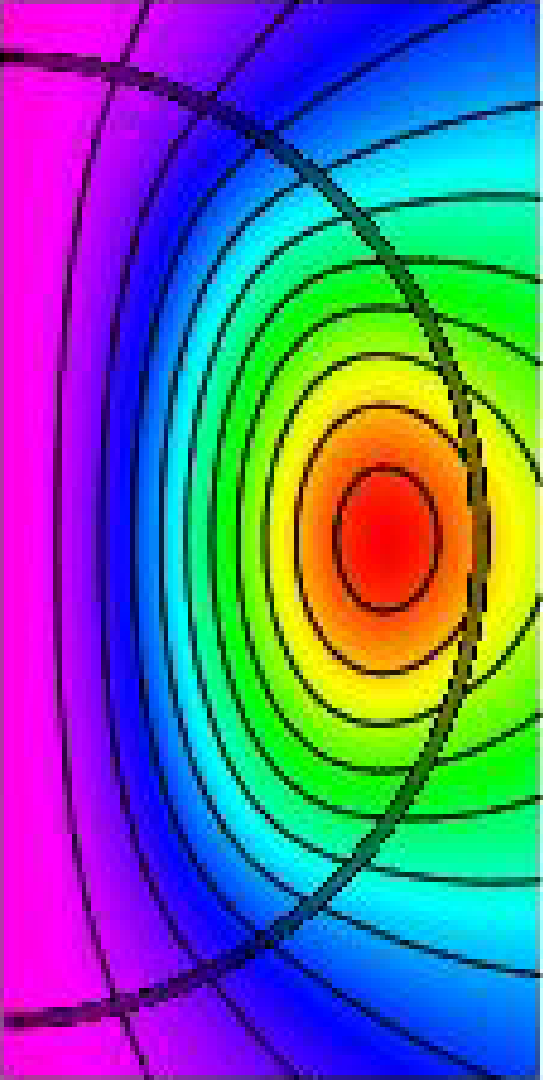}
b\includegraphics[width=0.45\columnwidth]{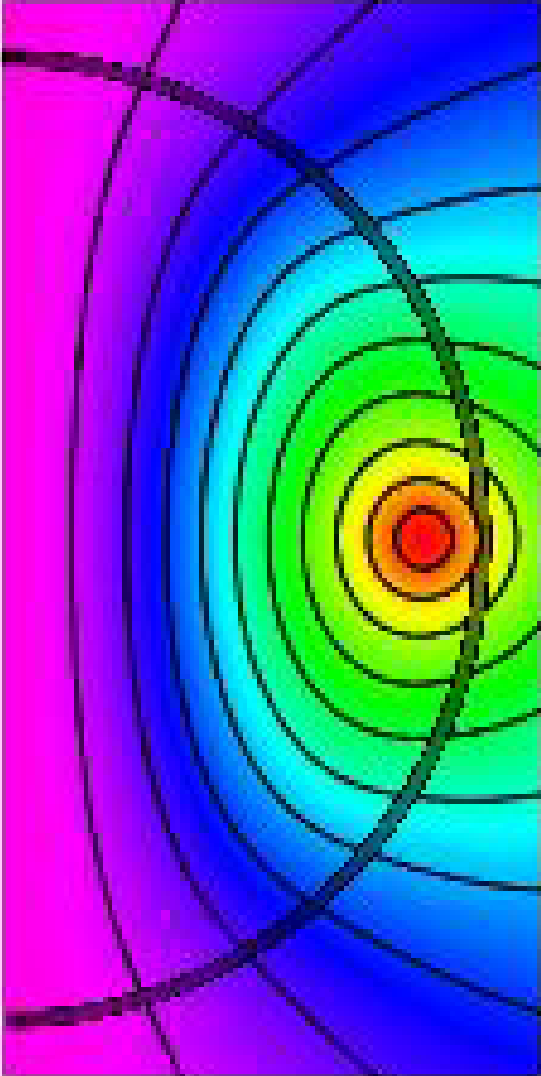}
\caption{The structure of the poloidal field lines with toroidal field as described in \S \ref{SOL_NUM}. In all cases we used $S_{\rm II}=S_{\rm III}=1$, $r_{in}=0$ $r_{*}=1$ and $n_{\rm e}=(r_{*}^{2}-r^{2})$. The colour represents the value of $\Psi$. Field structure for $\alpha=10$ (a) and  $\alpha=30$ (b). The toroidal field is hosted in the enclosed loops, which shrink as $\alpha$ increases. This solution was implemented in cylindrical geometry  $(R, z)$ with resolution $(400 \times 400)$.}
\label{Fig:8}
\end{figure}

We have solved the equation numerically, including a toroidal field in region III. The results presented in the figures and discussed correspond to $\zeta=1.1$. As can be seen in Fig.~\ref{Fig:8}, increasing the coefficient $\alpha$ leads to a progressive shrinkage of region III and to an increase of the maximum compared to a solution with the same parameters as in the purely poloidal field. The maximum of $\Psi$ occurs closer to the surface of the neutron star for higher $\alpha$. The maximum of the poloidal flux becomes higher as the system requires a stronger poloidal field to support the extra toroidal field. In the MHD barotropic equilibrium case, as the toroidal flux leads to a greater increase in the magnetic pressure, compared to the magnetic tension, the toroidal loops are pushed closer to the surface. The field now accommodates higher order multipoles of small amplitude, superimposed to the overall dipolar structure. This can be seen from the Grad-Shafranov equation, as the presence of the term $II'=\frac{1}{2}\frac{d I^{2}}{d\Psi}$ causes the maximum of the solution to rise and become more narrow, thus the loops hosting the toroidal field shrink. 

\begin{figure}
\centering
a\includegraphics[width=0.45\columnwidth]{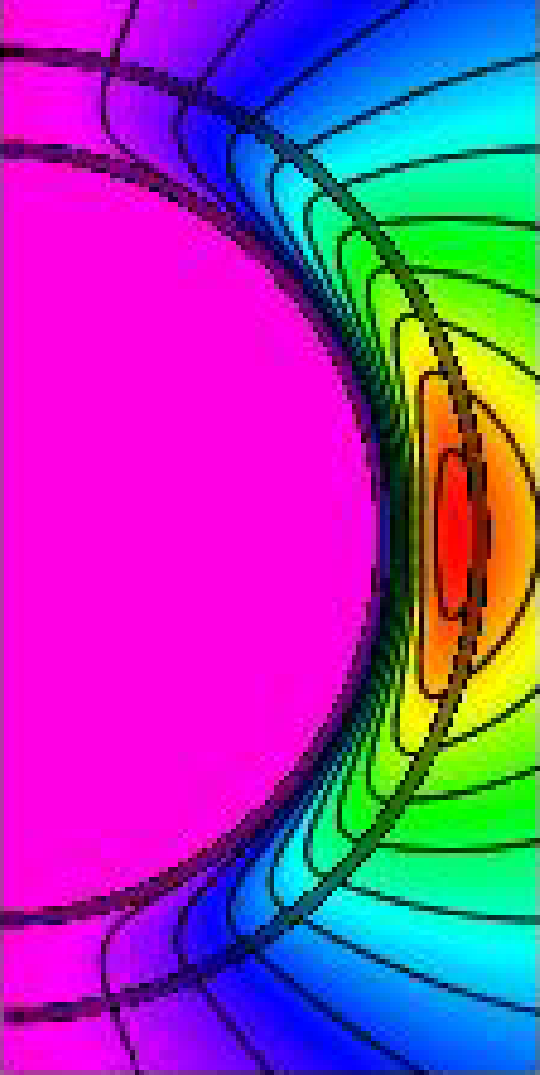}
b\includegraphics[width=0.45\columnwidth]{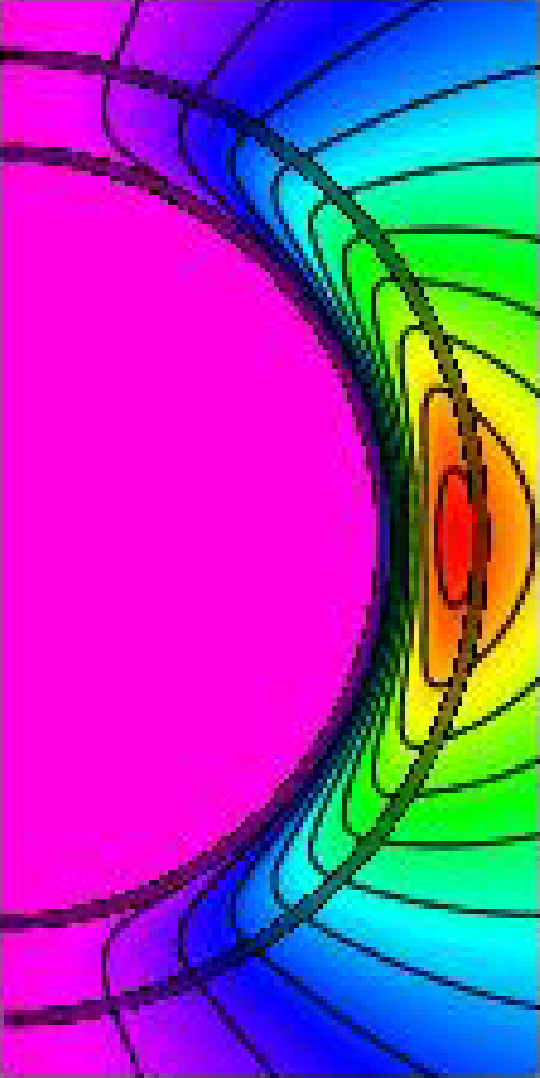}
\caption{The structure of the poloidal field lines with toroidal field as described in \S \ref{SOL_NUM}, in all cases we used $S_{\rm II}=S_{\rm III}=1$, $r_{in}=0.8$, $r_{*}=1$ and $n_{\rm e}=(r_{*}^{2}-r^{2})$. The colour represents the value of $\Psi$. Field structure for $\alpha=15$ (a) and $\alpha=45$ (b). The toroidal field is hosted in the enclosed red loops which shirk as $\alpha$ increases. This solution was implemented in cylindrical geometry  $(R, z)$ with resolution $(400 \times 400)$.}
\label{Fig:9}
\end{figure}

The energy content of the toroidal field is in general a small fraction of the total magnetic energy. We found that the toroidal energy is $\sim 3\%$ for $\alpha=20$, in the full sphere case, and $\sim 1\%$ in the crust solution for $\alpha=100$. Locally however the intensity of the toroidal field is a few times stronger than the poloidal. Analytical solutions of self-similar fields \citep{Lynden-BellBoily:1994, Gourgouliatos:2008, Beloborodov:2009, GourgouliatosVlahakis:2010} found that as the twist of the magnetic field lines increases, the toroidal component of the field tends to concentrate near the maximum of the flux function and eventually in an infinitesimally thin sheet or wire depending on the geometry of the system, while the energy carried by the toroidal field increases, reaches a maximum and then decreases. These effects have also been found in the numerical solutions of \cite{Lander:2009}, which were done in an MHD context. These effects on the behaviour of $\Psi$ are visualised in Figures \ref{Fig:4}  -- \ref{Fig:7}, where we plot the dependence of $\Psi$ on $r$ along the equator, and $\Psi$ on $\mu$ at $r_{\rm max}$ which is the radius where the maximum occurs. Sections of the poloidal flux function, showing the geometry of the poloidal field lines, are plotted in Figure \ref{Fig:8} for the full star and in Figure \ref{Fig:9} for the crust. We have not explored solutions for $\alpha$ above $35$ in the full sphere and $110$ for the crust because of numerical difficulties, thus we have not seen the decrease of the energy carried by the toroidal component of the field. As these solutions have $n_{\rm e}$ that decreases drastically from the centre to the surface, we find that the maximum is less steep than the one found in the example of uniform density, with the inhomogeneous term being smaller closer to the surface.  

As we explored the parameter space of $\zeta$, we find that for $\zeta>0.5$ the results are qualitatively similar to the ones discussed above, note however that for $\zeta<0.5$, $II'$ becomes infinite on the boundary of regions II and III. In general, for smaller $\zeta$ (but larger than 0.5) and larger $\alpha$ we can have stronger toroidal fields, but the energy carried by them is subdominant to the poloidal field energy. In particular, for $\alpha \sim 30$ and $\zeta$ in the range of $0.7$ to $1.5$ the toroidal field energy is only a few percent of the poloidal field energy. The mathematical form of equation (\ref{GS}-III) shows that for such choices of $\zeta$ we always add a positive quantity to the Grad-Shafranov operator, thus we are making the maximum sharper and steeper so that the toroidal magnetic field occupies a smaller volume. To expand the volume occupied by the toroidal magnetic field one needs to add a negative $II'$ term. To do so in this context of power law dependence we would need a negative $\zeta$, which is unacceptable, as it leads to infinite toroidal fields.

In the case where we solve only for the crust, we have explored a different density profile $n_{\rm e} \propto (r_{*}-r)^{4}$ which is a good approximation for the crust \citep{Cumming:2004}. This solution leads to a more extended region III, and a less steep maximum, as expected from the mathematical form of the Grad-Shafranov equation. 

We have also experimented with $S^{\prime}(\Psi) \propto \Psi$ and we found that the system converges to a structure that has a strong dipolar component and weaker multipoles as expected from our previous discussion. We remark however that this leads to differential rotation within the star, with the constraint that each field line rotates rigidly. 

We find consistent results when we apply a different code, the details of which will be presented in the future \citep{Armaza:2013}. This code solves iteratively a finite-difference version of the GS equation, starting from an initial seed until it converges. As a boundary condition, it assumes a multipolar expansion, consistent with equation (\ref{GS}-I), for the flux function outside the star with an arbitrary number of multipoles.  The coefficients of such an expansion are solved self-consistently with the inside solution by demanding continuity of $\Psi$ at the stellar surface. For the magnetic functions $I(\Psi)$ and $S(\Psi)$ considered here, with and without crust, the code converges to a predominantly dipolar solution, with negligible contributions of higher multipoles, in agreement with the results discussed above.

\subsection{Electron velocity profile}

Knowing the form of the magnetic field it is possible to evaluate the currents flowing in the crust. The electric current is due to the motion of the electrons, thus we can solve for the velocity profile. Taking the curl of equation (\ref{BFIELD}), in regions II and III the electric current is
\begin{eqnarray}
\bm{j} = \frac{c}{4 \pi}(\nabla I \times \nabla \phi - \Delta^{*} \Psi \nabla \phi)\,.
\end{eqnarray}
Substituting from equation (\ref{GS}) for the Grad-Shafranov operator, and given that the current is due to the electron motion, we obtain
\begin{eqnarray}
\bm{v_{p}}=-\frac{c}{4 \pi{\rm e} n_{\rm e}}I'\nabla \Psi \times \nabla \phi=-\frac{c}{4 \pi{\rm e}n_{\rm e}}I'\bm{B_{p}}\,,
\end{eqnarray}
\begin{eqnarray}
\bm{v_{\phi}}=-\frac{c}{4 \pi{\rm e} }\Big(\frac{II'}{r (1-\mu^2)^{1/2} n_{\rm e}} + r(1-\mu^{2})^{1/2}S'\Big)\bm{\hat{\phi}}\\ \nonumber
=-\frac{c}{4 \pi{\rm e} } \Big(I' \frac{\bm{B_{\phi}}}{n_{\rm e}} + r(1-\mu^{2})^{1/2}S'\bm{\hat{\phi}} \Big) \,.
\end{eqnarray}
We find that the velocity of the electron fluid has a component which is parallel to the magnetic field (force-free) and a second azimuthal component related to the scalar function $S'$. The angular velocity of the electron fluid after having subtracted any motion along the field lines, corresponding to the second component, is given by
\begin{eqnarray}
\Omega_{{\rm e}}=\frac{c}{4\pi {\rm e}}S'\,.
\label{OMEGA}
\end{eqnarray} 
In the examples, we have solved we have chosen $S^{\prime}=$const.~and we have found that the angular structure of the solution, in the absence of a toroidal field, is a dipole. To test the robustness of the inverse, we next modified the numerical scheme so that we assume a purely poloidal field that connects to a dipole field on the surface of the neutron star, but we allow $S^{\prime}$ to be free so that the solver chooses the appropriate value from the boundary conditions in a process similar to the simultaneous relaxation solution used in the solution for pulsar magnetospheres of \cite{Contopoulos:1999}. Having chosen a dipole field as the external solution, continuity of $\Psi$ requires $\Psi(r_{*},\mu)=\Psi_{0}(1-\mu^2)$. Substituting into equation (\ref{GS},II) we find
\begin{eqnarray}
S'=-\frac{1}{r_{*}^{2}n_{\rm e}(r_{*})}\Big(\frac{1}{1-\mu^{2}} \frac{\partial^{2} \Psi}{\partial r^{2}}\Big|_{r=r_{*}} -\frac{2}{r_{*}^{2}}\Big)\,.
\label{SPRIME}
\end{eqnarray}
Having evaluated the value of $S'$ on the boundary we can write it as a function of $\Psi$ and then proceed with the numerical solution inside the star, iterating until convergence. 

We find that the system indeed relaxes to a solution with $S'(\Psi)$ constant except for a small variation ($|\delta S^{\prime}(\Psi)/\bar{S}^{\prime}(\Psi)| <0.02$), compared to the average value $\bar{S}^{\prime}(\Psi)$. The average value found in this scheme is less than $5\%$ different from the value of the corresponding analytical solution. When we chose a quadrupole field as the boundary condition however, the system could not relax to an acceptable solution with a single valued function $S'(\Psi)$. Instead $S^\prime$ would have different values for two values of $\mu$ which nevertheless correspond to the same $\Psi$. 

We conclude that this is a strong indication that a dipole external field requires a uniformly rotating electron fluid to support it. The presence of toroidal field induces some higher order multipoles in the equilibrium solution, however the dipole field is always dominant.

\section{Discussion}

In this section, we discuss the application of our solutions to MHD equilibria (\S 4.1), evolution of magnetic fields in neutron star crusts (\S 4.2), and the possible interplay of MHD and Hall equilibria (\S 4.3).

\subsection{Application to barotropic equilibria}

As we discussed in \S 2, there is a direct mapping between solutions for Hall equilibria and MHD equilibria. The solutions found in this paper are based on the assumption that the density is a function of the radius only, which means that in the MHD case both the gravity and pressure gradient forces are radial and so they cannot formally balance the angular component of the magnetic force. This will lead to some deformation from spherical symmetry or ellipticity which has been neglected in our calculations. However, this ellipticity is small, being roughly equal to the ratio of the magnetic energy to the gravitational potential energy \citep{Ferraro:1954,Haskell:2008}, of the order of $10^{-7}$ for a neutron star with $B\sim10^{14}$G. In turn, the resulting perturbation to the magnetic field because of the difference in density profile is also small, so assuming the density is a function of radius only when deriving the field structure is in fact an excellent approximation. \cite{Akgun:2013} show that in the limit of a weak magnetic field compared to pressure and gravity, the magnetic equilibrium can be approached as a perturbation on the background.

Our solutions extending to the centre of the star are very similar to the final state of the magnetic evolution found by \cite{BraithwaiteNordlund:2006} and \cite{BraithwaiteSpruit:2006} who studied the evolution of tangled fields in non-barotropic stars towards a stable steady state and are in general accordance with other studies \citep{Ciolfi:2009, Duez:2010, LanderJones:2012}. For example, the choice of $I(\Psi)$ and $S(\Psi)$ in our numerical solutions with mixed toroidal and poloidal fields is the same as the ``type Ia'' solutions of \cite{LanderJones:2012} (see also eq.~[42] of \citealt{Duez:2010}). They also found other choices of $S(\Psi)$ that led to solutions with an external dipole magnetic field, such as $S^\prime(\Psi)\propto \Psi$ and a discontinuity in $S^\prime(\Psi)$ within the star (their type Ib and Ic solutions respectively, see eqs.~[14] and [15] of \citealt{LanderJones:2012}) which we have been able to reproduce, within the limit of weak fields.

Our numerical approach is different to previous work, which relies on a Green's function method to solve the Grad-Shafranov equation \citep{Hachisu:1986, Tomimura:2005,Lander:2009}. Instead, we apply the Gauss-Seidel method to the Grad-Shafranov equation directly. The fact that in the numerical treatment of the problem we have allowed the boundary between regions II and III to adjust itself is an improvement to previous studies of \cite{Lyutikov:2010}, and we have avoided the formation of current sheets on the surface of the star \citep{BroderickNarayan:2008}. The solutions confined in the crust have inevitable current sheets at the inner boundary of the crust, and such discontinuities can also be present in studies where the field penetrates the superconducting medium \citep{Lander:2012}.

\subsection{Evolution of crust fields due to the Hall effect and Ohmic dissipation}
\label{EVOLUTION}

Hall drift has generally been discussed as leading to enhanced magnetic dissipation in neutron star crusts. \cite{GoldreichReisenegger:1992} proposed that it would lead to a cascade of magnetic energy to small scales where it would dissipate Ohmically. Numerical simulations in cartesian boxes do indeed show this and have clarified the scalings of the turbulent cascade \citep{Biskamp:1996, Cho:2009, Wareing:2010}. Alternatively, in magnetars, Hall drift has been discussed as providing an efficient transport of helicity into the magnetosphere, where it can drive flaring behavior \citep{Thompson:1995}.

In contrast, simulations of the field evolution in spherical geometry show that, while there is rapid evolution initially, the Hall drift quickly saturates and the field evolves on a slower Ohmic timescale \citep{PonsGeppert:2010, Kojima:2012, Vigano:2012, Vigano:2013}. These simulations so far have assumed axisymmetric fields, which could be responsible for the different outcome. However, the fact that there is a family of steady-state solutions for the Hall effect in the crust suggests an alternative picture in which the field initially evolves rapidly due to the Hall effect because the initial condition is far from equilibrium, but as the field approaches an equilibrium state, the Hall drift slows down. \cite{PonsGeppert:2007} noted that after the fast initial transient evolution, the field in their simulation reaches a ``quasi-equilibrium field'' with a slow decay. 

Whether the simulations are evolving into a Hall equilibrium state is not clear, but assuming this to be the case, we can make a simple model of the field evolution by assuming that the Hall drift is rapid enough that the field structure and geometry remains that of a Hall equilibrium as Ohmic decay operates and reduces the overall magnetic energy. Note that Ohmic dissipation will preserve the angular structure of the field provided that the diffusivity only depends on radius, as dipoles and higher order multiples are angular eigenstates of the dissipation operator and they do not couple. 

Consider the poloidal dipole Hall equilibrium discussed in \S \ref{POLOIDAL} for $n_{\rm e}=$const.~ and $r_{in}=0$. In that solution, the electron fluid is rigidly rotating, and assuming that the Hall term acts to maintain rigid rotation as the currents Ohmically decay, the effect is to reduce the angular velocity of the electrons with time. To calculate the braking rate, we write down an equation for the rate of change of magnetic energy due to Ohmic dissipation. We first write $\Psi$ in terms of $\Omega_{\rm e}$ using equation (\ref{OMEGA}),
\begin{eqnarray}
\Psi=(1-\mu^{2})\frac{2\pi {\rm e} \Omega_{\rm e} n_{\rm e}}{c}\Big(\frac{r^{2}}{3}-\frac{r^{4}}{5}\Big)\,.
\end{eqnarray}
Assuming the global structure to be preserved, we then substitute this form for $\Psi$ into
\begin{eqnarray}
\frac{d}{dt}\int\frac{1}{8 \pi} \bm{B}^{2}dV= -\int \frac{1}{\sigma} \bm{j}^{2}dV\,.
\end{eqnarray}
which gives
\begin{eqnarray}
\frac{d \Omega_{\rm e}}{dt}=-\frac{3.42 c^{2}}{\pi \sigma r_{*}^{2}}\Omega_{\rm e}\,,
\end{eqnarray}
for $\sigma(r)=$const. We find that $\Omega_{\rm e}$ drops exponentially with time, slowing down the electron fluid within the crust on a timescale that is faster than the slowest Ohmic dissipation rate for a field that connects to an external dipole which is ${3.09c^{2}}/{\pi \sigma r_{*}^{2}}$. By keeping the electron fluid close to rigid rotation, the Hall term acts to maintain the amplitudes of higher order Ohmic modes, enhancing the dissipation rate compared to pure Ohmic decay.

We stress that while transient evolution to a Hall equilibrium state qualitatively matches the behavior seen in numerical simulations, further work is needed to investigate whether the field is actually evolving in this way. Indeed, it is important to emphasise that it is not even clear whether a magnetic field would naturally evolve to equilibrium under Hall drift, as the equilibria are not necessarily attractors in electron MHD. This is because the energy principle applicable to MHD is not appropriate for Hall evolution. Whereas equilibrium solutions in MHD map directly to Hall equilibria, the dynamics is quite different. It will be of great interest to look in detail at the  field structures and currents that emerge in the time-dependent calculations to determine whether the slowly evolving field observed in simulations is close to a Hall equilibrium state.

\subsection{General implications for neutron star magnetic field evolution}

Neutron star evolution potentially involves an interesting interplay between MHD and Hall equilibria, as the solidification of the neutron star crust occurs after many Alfven crossing times \citep{Ciolfi:2009}. Therefore, the field structure is presumably in an MHD equilibrium state when the crust forms and starts to obey Hall dynamics. 

For example, an important difference between MHD and Hall equilibria could be their stability. In this paper, we have calculated equilibria without addressing whether they are stable or not. One possibility is that, whereas twisted-torus type MHD equilibria are stable \citep{BraithwaiteSpruit:2004}, the corresponding equlibria in electron MHD may not be stable. \cite{Cumming:2004} found that the purely poloidal field is neutrally stable under Hall evolution, but the stability of mixed poloidal-toroidal fields has not been addressed. If there was such a difference, this could be a driver of magnetic activity in young neutron stars as the crust forms and the previously stable MHD equilibrium field would become unstable under the action of Hall drift. In general, steep derivatives of the electron velocity excite instabilities as it has been shown for plane parallel geometry \citep{RheinhardtGeppert:2002, PonsGeppert:2010}, however, spherical geometry instabilities are more complicated \citep{Reisenegger:2007} as non axially symmetric modes may become unstable.

Another important effect could be the difference between confined fields, which can have a large fraction of energy in a toroidal component, and fields that are not confined, for which only a small fraction of the energy can lie in the toroidal field. Newborn neutron stars are believed to host strong toroidal fields and higher order multipoles \citep{Rea:2010, ShabaltasLai:2012}  which are buried inside the neutron star. We have provided a solution that is fully confined within the neutron star, \S \ref{CONFINED}, which is an extreme example of this configuration. If the neutron star starts with a mostly confined magnetic field with poloidal and toroidal magnetic components, the two components could contain about the same amount of energy \citep{Ciolfi:2009}. As we have shown, for fields that are not fully confined in the star, only a small fraction of energy can be hosted by the toroidal field, at least for axially symmetric equilibria. This difference requires that older neutron stars should have undergone a stage when they were expelling toroidal fields.

Even under the approximation of the barotropic solutions as an initial condition, the magnetic field will not be in Hall equilibrium as the electron number density $n_{\rm e}$ is different from the mass density $\rho$, because the number of electrons per unit mass $Y_{\rm e}\equiv n_{\rm e}/\rho $ varies with $r$. Assuming a magnetic field that satisfies equation (\ref{BAR2}) we substitute into equation (\ref{BEVOL}). In the Hall description we shall use the electron density $n_{\rm e}=Y_{\rm e}\rho$; leading to the evolution of the magnetic field by
\begin{eqnarray}
\delta \bm{B} = -\frac{c}{4 \pi {\rm e}} \nabla \frac{1}{Y_{\rm e}}\times \nabla S_{B}\delta t\,.
\end{eqnarray}
$Y_{\rm e}$ is a function with a very steep gradient near the neutron drip ($4\times 10^{11}$g~cm$^{-3}$). We are going to focus on this point assuming it occurs at some $r=r_{d}$, thus $\nabla Y_{\rm e} =\frac{d Y_{\rm e}}{dr}\bm{\hat{r}}$. Using as an example the dipole solutions $S_{B}= g(r)\sin^{2}\theta$, we find
\begin{eqnarray}
\delta {\bm{B}} =-\frac{c}{2 \pi {\rm e}} \frac{g(r)}{r Y_{\rm e}^{2} } \frac{dY_{\rm e}}{dr}\sin\theta\cos\theta \delta t \bm{\hat{\phi}}\,.
\end{eqnarray}
Thus this will  spontaneously generate a quadrupole toroidal field inside the crust. Depending on the steepness of the gradient $Y_{\rm e}$, this field may be concentrated in a thin shell and dissipate fast, providing a source of thermal energy. However, the detailed evolution from this point onwards is beyond the capacity of the numerical scheme presented in this paper and requires the use of a simulation. The generation of this toroidal field, however, demonstrates that switching to a solid lattice changes the equilibrium requirements from a barotropic MHD fluid state. It also provides a spontaneous mechanism for the generation of higher order multipoles which have been discussed in the context of magnetar activity and rejuvenate the energy supply. 

A general picture for field evolution is then that, starting from an MHD equilibrium, after some Hall timescale ($\sim {10^{5}}/{B_{15}}$~years, where $B_{15}={B}/{10^{15}G}$), the field relaxes to a stable Hall equilibrium. The difference between MHD and Hall equilibria drives the expulsion of toroidal loops of magnetic field which power flaring activity \citep{Thompson:1995,BraithwaiteSpruit:2006}. Even when the magnetic field in the crust reaches a stable Hall equilibrium, it  will Ohmically decay and continue to exert stresses on the lattice as $\bm{j}\times \bm{B}\neq0$. Strong fields ($B\sim10^{15}$G) are known to exceed the maximum strain of the crust \citep{Thompson:1995}, but as is evident from the solutions presented in this paper, the field inside the crust can be an order of magnitude stronger compared to the inferred dipole value. Such hidden strong fields are consistent with magnetar activity from not so strongly magnetised neutron stars \citep{Dall'Osso:2012}. As we have stressed already, it is important to confront the calculations of Grad-Shafranov equilibria and numerical simulations of crustal fields to investigate further the role of Hall equilibria in neutron star field evolution.

\section{Conclusions}

In this paper, we have investigated Hall equilibria for magnetic fields. Such equilibria are given as solutions to the Grad-Shafranov equation, which is a well studied mathematical equation. Analytical solutions known in other astrophysical contexts are indeed applicable to the context of Hall equilibria, whereas we have solved numerically the Grad-Shafranov equation in cases where a toroidal field was confined in regions of closed poloidal field lines inside the star. We find that our numerical scheme reproduces the known analytical solutions. We stress the importance of solutions which are connected to a dipole external field: these solutions correspond to uniformly rotating electron fluids. The inclusion of higher order multipoles is likely to be associated with velocity gradients in the electron fluid which lead to evolution. Hall simulations focus more on the evolution of the magnetic field and its Ohmic dissipation and are not easily comparable with our solutions, nevertheless, our solutions bear little difference with the relaxed states that dissipate Ohmically found through these simulations, especially in the fact that the field is dominated by a dipole, it retains its shape, and its intensity decreases exponentially with time. We stress the importance of these results in the context of magnetars. Even if a young magnetar comes from a barotropic progenitor in magnetic equilibrium, it will not be in Hall equilibrium as the electron density does not directly reflect the mass density. This excites higher order multipoles which may be associated to magnetar activity. It may be plausible that through this activity the field relaxes to a Hall equilibrium that dissipates Ohmically.

\section*{Acknowledgements}  
KNG is supported by the Centre de Recherche en Astrophysique du Qu\'ebec. AC is supported by an NSERC Discovery Grant and is an Associate Member of the CIFAR Cosmology and Gravity Program. AR and JAV are supported by FONDECYT Regular Grants 1110213 and 1110135, respectively. CA is supported by a CONICYT Master's Fellowship.
AR, CA, and JAV are supported by CONICYT International Collaboration Grant DFG-06. AR and CA are supported by the Basal Center for Astrophysics and Associated Technologies. We thank Pablo Marchant for his comments on Hall stability. 

\bibliographystyle{mnras}
\bibliography{BibTex.bib}

\section*{Appendix}

We have implemented the Gauss-Seidel method in Fortran for the solution of the Grad-Shafranov equation (\ref{GS}), both in a cylindrical and a spherical grid. We start with a trial solution, in every iteration we evaluate the Grad-Shafranov equation and we correct accordingly the trial solution until convergence. Here we present the solution in spherical geometry. The grid points are numbered with $i,j$, where $i$ is related to the radial distance from the centre running from $i_{1}=0$ at $r_{1}=0$ to $i_{2}$ at $r_{2}=10$, while $dr=(r_{2}-r_{1})/(i_{2}-i_{1})$. We have normalized the radius of the star to $r_{*}=1$ and $i_{*}=\|r_{*}/dr\|$ is given by rounding to the nearest integer. The cosine of the polar angle $\mu$ is given by the variation of $j$'s, so that the south pole $\mu=-1$ is at $j_{1}$, the north pole $\mu=1$ is at $j_{2}=-j_{1}$, and the equator $\mu=0$ corresponds to $j=0$, while $d\mu=2/(j_{2}-j_{1})$, thus a point $i,j$ on the grid corresponds to a point whose spherical coordinates are $r=idr$ and $\mu=jd\mu$. The $n^{th}$ iteration of $\Psi$ at point $i,j$ will be given by
\begin{eqnarray}
\Psi_{i,j}^{n}=\Psi_{i,j}^{n-1}+ L_{i,j}^{n}\,.
\end{eqnarray}
Whether a given point lies in region II or III is determined in each iteration by comparing the local value $\Psi_{i,j}$ with the value at the equator, $\Psi_{i_*,0}$. This translates into 
\begin{eqnarray}
L_{i,j}^{n}=h\Big( (idr)^2\frac{\Psi_{i-1,j}^{n}+\Psi_{i+1,j}^{n-1}-2\Psi_{i,j}^{n-1}}{dr^{2}} \nonumber \\
+(1-(jd\mu)^{2})\frac{ \Psi_{i,j-1}^{n}+\Psi_{i,j+1}^{n-1}-2\Psi_{i,j}^{n-1}}{d\mu^{2}} \nonumber \\ 
+ S'(idr)^{4}n_{e~i,j}(1-(jd\mu)^2) \nonumber \\
+ H(\Psi_{i,j}^{n-1} -\Psi_{0})1.1 \alpha^{2} (idr)^2\ \nonumber \\
\times (\Psi_{i,j}^{n-1}-\Psi_{0})^{1.2}\Big) \frac{1}{(idr)^2}\,,
\end{eqnarray}
where $H$ is the Heaviside step function so that the solver distinguishes between regions II and III. We remark that we use the updated values of $\Psi$ whenever possible, which accelerates the convergence, while we choose $h=\frac{1}{50} dr d\mu$, which increases the number of iterations needed to reach the solution but reduces the chance of divergence; formally it must be $h\leq \frac{1}{4}dr d\mu$. The boundary conditions chosen are $\Psi_{i_{1},j}=\Psi_{i_{2}, j}=\Psi_{i, j_{1}}=\Psi_{i, j_{2}}= 0$, and we repeat the iteration while $i_{1}<i<i_{2}$ and $j_{1}<j<j_{2}$. As an alternative, we also used a dipole field at the $i_{2}$ boundary of the domain normalized at every iteration by the value we found at the equator of the star $\Psi_{i_{2}, j}= \Psi_{i_{*},0} (1- (j d\mu)^2)/(i_{2}dr)$. We found that in both choices of boundary conditions the solution inside the star was very weakly affected. We used this scheme with a chosen value for $S'$ and $\alpha$, i.e. in the dipole constant density solution, Figures \ref{Fig:4} -- \ref{Fig:9}, we chose $S'=1$ and $\alpha$ was $(0, 10, 20, 30)$ for the full star and $(0, 50, 100)$ for the $r_{in}=0.8$ crust. We also implemented a varying grid in $R$ where $R=r/(r+1)$, $0\leq R \leq 1$, so that $R=1$ corresponds to infinity. In that case we have set $\Psi(R=1)=0$ and we found indistinguishable results compared to the previous grid.

\label{lastpage}

\end{document}